\documentclass[lettersize,journal]{IEEEtran}
\usepackage[caption=false,font=scriptsize,labelfont=normalfont,textfont=normalfont]{subfig}
\usepackage[normalem]{ulem}
\usepackage{algorithmic,algorithm}
\usepackage[tbtags]{amsmath}
\usepackage{amsfonts,amssymb}
\usepackage{array}
\usepackage{bm,bbm,booktabs,pifont}
\usepackage{calc}
\usepackage{cite}
\usepackage{graphicx}
\usepackage{graphbox} %loads graphicx package
\usepackage{hhline}
\usepackage[usenames,dvipsnames]{xcolor}
\usepackage{colortbl}
\usepackage{rotating}
\usepackage{lipsum}
\usepackage{makecell}
\usepackage{mathtools}
\usepackage{microtype}
\usepackage{multirow}
\usepackage{numprint}
\usepackage[free-standing-units,binary-units]{siunitx}
\usepackage{stfloats}
\usepackage{tabularx}
\usepackage{textcomp}
\usepackage{tikz}
\usepackage{verbatim}
\usepackage{footnote}
\usepackage{makecell}

\definecolor{tab_blue}{HTML}{1F77B4}
\definecolor{tab_orange}{HTML}{FF7F0E}
\definecolor{tab_green}{HTML}{2CA02C}
\definecolor{tab_red}{HTML}{D62728}
\definecolor{tab_purple}{HTML}{9467BD}
\definecolor{tab_brown}{HTML}{8C564B}
\definecolor{tab_pink}{HTML}{E377C2}
\definecolor{tab_gray}{HTML}{7F7F7F}
\definecolor{tab_olive}{HTML}{BCBD22}
\definecolor{tab_cyan}{HTML}{17BECF}

\newcommand{\change}[1]{#1}

\makeatletter
\newcommand*{\inlineequation}[2][]{%
  \begingroup
    \refstepcounter{equation}%
    \ifx\\#1\\%
    \else
      \label{#1}%
    \fi
    \relpenalty=10000 %
    \binoppenalty=10000 %
    \ensuremath{%
       \displaystyle % larger fractions, ...
      #2%
    }%
    ~\hfill\@eqnnum
  \endgroup
}
\makeatother

\input{def.set} % for \argmin
\DeclarePairedDelimiterX{\distx}[2]{(}{)}{%
  #1\;\delimsize\|\;#2%
}
\newcommand{\dist}{\mathcal{E}\distx}
\newcommand{\distDP}{\hspace{1pt}\overline{\operatorname{SegSNR}}}

\DeclarePairedDelimiter\ceil{\lceil}{\rceil}

\newcolumntype{L}{>{\raggedright\arraybackslash}X}
\newcolumntype{R}{>{\raggedleft\arraybackslash}X}
\newcolumntype{M}[1]{>{\raggedright\arraybackslash}m{#1}}

\DeclareSIUnit{\spkr}{spkr}
\DeclareSIUnit{\nothing}{\relax}

\def\useurl{1}
\if\useurl1
  \usepackage{hyperref}
  \newcommand\myshade{85}
  \colorlet{mylinkcolor}{violet}
  \colorlet{mycitecolor}{YellowOrange}
  \colorlet{myurlcolor}{Aquamarine}
  \hypersetup{
    linkcolor  = mylinkcolor!\myshade!black,
    citecolor  = mycitecolor!\myshade!black,
    urlcolor   = myurlcolor!\myshade!black,
    colorlinks = true,
  }
\else
  \usepackage{url}
\fi

\newcommand{\SSfull}{\mathbb{S}}
\newcommand{\SSnpretrain}{\mathbb{\tilde{S}_\text{\textbf{p}-tr}}}

\newcommand{\SSpretrain}{\mathbb{S_\text{\textbf{p}-tr}}}
\newcommand{\SSpreval}{\mathbb{S_\text{\textbf{p}-vl}}}
\newcommand{\SStrain}{\mathbb{S_\text{\textbf{f}-tr}}}
\newcommand{\SSval}{\mathbb{S_\text{\textbf{f}-vl}}}
\newcommand{\SStest}{\mathbb{S_\text{te}}}
\newcommand{\SAfull}{\mathbb{G}}
\newcommand{\SAtrain}{\mathbb{G_\text{tr}}}
\newcommand{\SAval}{\mathbb{G_\text{vl}}}
\newcommand{\Mfull}{\mathbb{M}}
\newcommand{\Mtrain}{\mathbb{M_\text{tr}}}
\newcommand{\Mval}{\mathbb{M_\text{vl}}}

\newcommand{\Nfull}{\mathbb{N}}
\newcommand{\Ntrain}{\mathbb{N_\text{tr}}}
\newcommand{\Nval}{\mathbb{N_\text{vl}}}
\newcommand{\Ntest}{\mathbb{N_\text{te}}}

\ExplSyntaxOn
\int_new:N \l__nebu_min_prefix_int
\tl_new:N \l__nebu_base_number_tl
\tl_new:N \l__nebu_mode_tl
\cs_new:Npn \nebu_prefix:n #1
  {
    \exp_args:Nf \__nebu_prefix:n
      { \exp_args:Nf \fp_to_scientific:n { \tl_lower_case:n {#1} } }
  }
\cs_new:Npn \__nebu_prefix:n #1
  { \__nebu_prefix:nwnw #1 \q_stop }
\cs_new:Npn \__nebu_prefix:nwnw #1 e #2 \q_stop
  { \__nebu_find_prefix:nnn {#2} {#1} {1} }
\cs_new:Npn \__nebu_find_prefix:nnn #1 #2 #3
  {
    \tl_if_exist:cT { l__nebu_ #1 \tl_use:N \l__nebu_mode_tl _prefix_tl }
      {
        \use_i_delimit_by_q_stop:nw
          { \__nebu_output:nnn {#2} {#1} {#3} }
      }
    \int_compare:nNnT {#1} < \l__nebu_min_prefix_int
      {
        \use_i_delimit_by_q_stop:nw
          {
            \exp_args:Nf \__nebu_find_prefix:nnn
              { \int_eval:n { \l__nebu_min_prefix_int } } {#2}
              { #3 / 1\prg_replicate:nn { \l__nebu_min_prefix_int - #1 }{ 0 } }
          }
      }
    \use_i:nn
      {
        \exp_args:Nf \__nebu_find_prefix:nnn
          { \int_eval:n {#1-1} } {#2} {#3*10}
      }
      \q_stop
  }
\exp_args_generate:n { fv }
\cs_new:Npn \__nebu_output:nnn #1 #2 #3
  {
    \exp_args:Nfv \nebu_output:nn
      { \fp_to_decimal:n {#1*#3} } { l__nebu_ #2 \tl_use:N \l__nebu_mode_tl _prefix_tl }
  }
\cs_new_protected:Npn \nebu_prefix_set:Nnn #1 #2 #3
  {
    \exp_args:Nxx \__nebu_prefix_set:nnN
      { \tl_trim_spaces:n {#2} } { \int_eval:n {#3} } #1
  }
\cs_new_protected:Npn \__nebu_prefix_set:nnN #1 #2 #3
  {
    \tl_clear_new:c { l__nebu_ #2 _prefix_tl }
    \tl_clear_new:c { l__nebu_ #2 _siunitx_prefix_tl }
    \tl_set:cn { l__nebu_ #2 _prefix_tl } {#1}
    \tl_set:cn { l__nebu_ #2 _siunitx_prefix_tl } {#3}
    \int_set:Nn \l__nebu_min_prefix_int { \int_min:nn {#2} { \l__nebu_min_prefix_int } }
  }
\NewExpandableDocumentCommand \prefix { m }
  { \nebu_prefix:n {#1} }
\cs_new:Npn \nebu_output:nn #1 #2 { #1\, \textrm{#2} }
\NewDocumentCommand \setprefix { m m m }
  { \nebu_prefix_set:Nnn #1 {#2} {#3} }
\DeclareSIPrefix \none { } { 0 }
\setprefix \none { } { 0 }
\cs_new_protected:Npn \__nebu_store:nn #1 #2
  {
    \tl_set:Nn \l__nebu_base_number_tl {#1}
    \cs_set:Npn \prefix {#2}
  }
\NewExpandableDocumentCommand \prefixSI { o m o m }
  {
    \group_begin:
      \cs_set_eq:NN \nebu_output:nn \__nebu_store:nn
      \tl_set:Nn \l__nebu_mode_tl { _siunitx }
      \nebu_prefix:n {#2}
      \SI [#1] { \l__nebu_base_number_tl } [#3] {#4}
    \group_end:
  }
\ExplSyntaxOff

\setprefix \yocto { y } { -24 }
\setprefix \zepto { z } { -21 }
\setprefix \atto  { a } { -18 }
\setprefix \femto { f } { -15 }
\setprefix \pico  { p } { -12 }
\setprefix \nano  { n } { -9 }
\setprefix \micro { \SIUnitSymbolMicro } { -6 }
\setprefix \milli { m } { -3 }
\setprefix \centi { c } { -2 }
\setprefix \deci  { d } { -1 }
\setprefix \deca  { da } { 1 }
\setprefix \hecto { h }  { 2 }
\setprefix \kilo  { k }  { 3 }
\setprefix \mega  { M }  { 6 }
\setprefix \giga  { G }  { 9 }
\setprefix \tera  { T }  { 12 }
\setprefix \peta  { P }  { 15 }
\setprefix \exa   { E }  { 18 }
\setprefix \zetta { Z }  { 21 }
\setprefix \yotta { Y }  { 24 }
\begin{document}

\title{
%  Improving Data and Resource Efficiency for Personalized Speech Enhancement
%  through Self-Supervised Learning
Efficient Personalized Speech Enhancement\\through Self-Supervised Learning
% Efficient Personalized Speech Enhancement
% \\through Self-Supervised Noisy Target Training
}

\author{
 Aswin Sivaraman,~\IEEEmembership{Student Member,~IEEE},
 Minje Kim,~\IEEEmembership{Senior Member,~IEEE}
}%

\markboth{IEEE Journal of Selected Topics in Signal Processing,~Vol.~X, No.~Y, ZZZ~2022}%
{Shell \MakeLowercase{\textit{et al.}}: A Sample Article Using IEEEtran.cls for IEEE Journals}

% \IEEEpubid{0000--0000/00\$00.00~\copyright~2021 IEEE}
% Remember, if you use this you must call \IEEEpubidadjcol in the second
% column for its text to clear the IEEEpubid mark.

\maketitle

\begin{abstract}
This work presents self-supervised learning methods for developing monaural speaker-specific (i.e., personalized) speech enhancement models. While generalist models must broadly address many speakers, specialist models can adapt their enhancement function towards a particular speaker's voice, expecting to solve a narrower problem. Hence, specialists are capable of achieving more optimal performance in addition to reducing computational complexity. However, naive personalization methods can require clean speech from the target user, which is inconvenient to acquire, e.g., due to \change{subpar recording conditions}. To this end, we pose personalization as either a zero-shot task, in which no additional clean speech of the target speaker is used for training, or a few-shot learning task, in which the goal is to minimize the duration of the clean speech used for transfer learning. With this paper, we propose self-supervised learning methods as a solution to both zero- and few-shot personalization tasks. The proposed methods are designed to learn the personalized speech features from unlabeled data (i.e., in-the-wild noisy recordings from the target user) without knowing the corresponding clean sources. Our experiments investigate three different self-supervised learning mechanisms. We set up a pseudo speech enhancement problem as a pretext task, which pretrains the models to estimate noisy speech as if it were the clean target. Contrastive learning and data purification methods regularize the loss function of the pseudo enhancement problem, overcoming the limitations of learning from unlabeled data. We assess our methods by personalizing the well-known ConvTasNet architecture to twenty different target speakers. The results show that self-supervised models achieve zero-shot and few-shot personalization using fewer model parameters and less clean data from the target user, achieving the data efficiency and model compression goals.
\end{abstract}

\begin{IEEEkeywords}
Personalized speech enhancement, self-supervised learning, data efficiency, model complexity
\end{IEEEkeywords}

\section{Introduction}\label{sec:intro}
\begin{figure}[t]
    \centering
    \subfloat[]{
      \includegraphics[width=0.45\columnwidth]{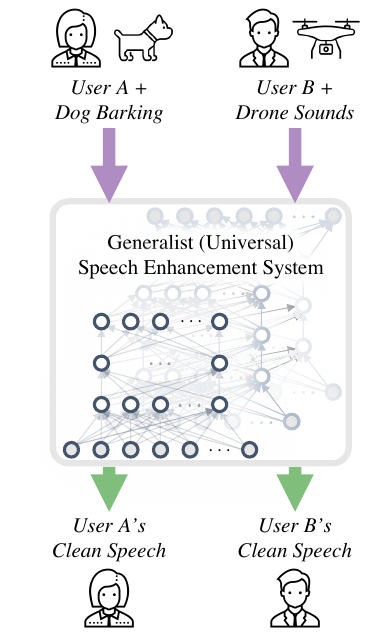}% 
      \label{fig:overview_generalist}}
    \hfill
    \subfloat[]{
      \includegraphics[width=0.45\columnwidth]{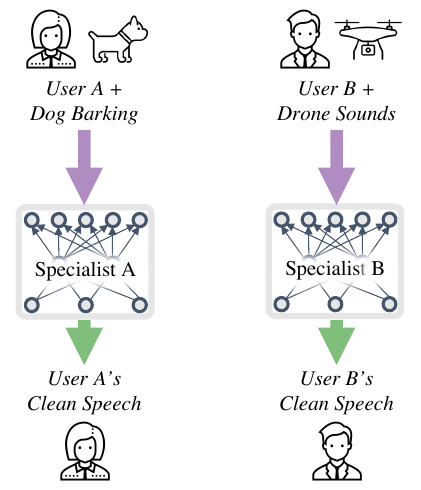}% 
      \label{fig:overview_specialist}}
    \caption{\change{A conceptual overview of how specialist models can replace generalist models for speech enhancement. Model inputs and outputs are shown in purple and green arrows, respectively. For example, Specialist A can train using User A's unique noisy speech recordings, producing a speaker-specific cleaned output. Optimizing for a specific speaker or environment can yield improved performance. Additionally, because specialists address a subset of the generalist's task, they can theoretically afford a reduction in model parameters.}}
    \label{fig:concepts}
\end{figure}

\IEEEPARstart{W}{ith} the ubiquity of voice-controlled intelligent devices, there is now an ever-growing demand for low-cost robust speech processing systems.
These systems are reliant on speech enhancement (SE) technology, which improves the quality and intelligibility of noisy speech signals \cite{TaherianH2020ieeeacmaslp}.
Over the last decade, deep learning algorithms have quickly defined the state-of-the-art in SE research \cite{HintonG2012ieeespm, XuY2014ieeespl, HuangP2015ieeeacmaslp, WeningerF2015lvaica, HersheyJ2016icassp, PascualS2017segan, ChoiHS2018phase-aware, TanK2018interspeech, LuoY2019conv-tasnet}.
Most neural networks proposed for SE are trained using supervised learning frameworks \cite{WangDL2018ieeeacmaslp}.
Typically, input-output pairs are programmatically generated by mixing various speech recordings with assorted noise recordings.
Supervised data preparation requires labeled datasets, i.e., the speech signals are known to be clean or of reference quality.
The neural networks then learn a mapping function between the input mixture signals and their originating ground-truth clean speech signal.
Consequently, the learning outcomes of supervised SE models are highly dependent on the diversity of the training data and increased model complexity.
But because large publicly available datasets do not represent all populations, machine learning models can easily become biased towards over-represented social groups \cite{MeyerJ2020biascorpus}.
For instance, a general-purpose universal SE model may under-perform for a particular target speaker if their unique vocal characteristics or noisy environment are never encountered during training.
And although modern GPU technology enables training neural networks with millions of learnable parameters \cite{DengL2014deeplearning}, bigger models are infeasible to deploy on low-resource devices \cite{ZhouZ2019edgecomputing}.
To overcome these disadvantages of generalist models, we focus our investigation on developing speaker-specific specialist models.
Through model specialization, we narrow down the scope of the SE task, unlocking the potential for improved enhancement of a single target speaker while affording a reduction in model complexity.
We refer to this narrowed problem definition as the personalized speech enhancement (PSE) task.
Figure \ref{fig:concepts} illustrates this idea of replacing large generalist models with smaller more-optimized specialist models.

While specialists are theoretically preferable to generalists, in practice, personalization is a challenging optimization strategy because it requires explicit knowledge about the target speaker and their environment.
Realistically, users of voice-controlled devices would be reluctant to share their clean voice data due to privacy concerns.
Given that speech synthesis models can be conditioned on speaker identity using only five seconds of clean speech data \cite{JiaY2018neurips}, unwanted forgery of one's voice is a genuine issue.
Even if users are willing to provide their clean speech, it can still be difficult to collect a large quantity of studio-quality noise-free recordings.
Users might not have access to a quiet anechoic room, and their microphones may introduce unwanted artifacts.
% Because background noise might always be present, recordings captured by the voice-controlled device must be treated as unlabeled data.

This study addresses the voice-controlled device PSE task considering the aforementioned user data constraints.
We envision two possible scenarios.
In one, the smart device must personalize using only the target speaker's in-the-wild noisy speech data.
We view this as a \textit{zero-shot} (ZSL) machine learning problem as there is no labeled data.
In the second scenario, the target user provides the device with a limited amount of clean speech, on the order of seconds.
If these signals are of good quality, we consider personalization to be a \textit{few-shot} (FSL) problem, given that some labeled speech is now available.
Even if there is abundant unlabeled data, the supervised training framework cannot be applied because the data is not guaranteed to be reference-quality.

In this paper, we assess two self-supervised learning (SSL) frameworks for developing PSE models.
Through SSL, we train the PSE model to solve a ``pretext task'' using the unlabeled data.
%---in-the-wild noisy speech, in our case.
The effectiveness of the model's learned features in SSL depends on how mismatched the pretext task is from the intended downstream task.
This paradigm has gained popularity in many research areas such as computer vision \cite{VondrickC2018eccv}, natural language processing \cite{DevlinJ2019bert}, and reinforcement learning \cite{JangE2018pmlr}.
One particular SSL technique, known as contrastive learning, augments the unlabelled data in a pairwise manner.
Training a model to keep similar samples together becomes an easy method for developing discriminative feature spaces.
As suggested in the recent SimCLR paper, the composition of the pairwise data augmentation pipeline strongly influences the robustness of the learned contrastive features \cite{ChenT2020simclr}.
\change{Principally, self-supervision differs from traditional supervised learning by removing the necessity for labeled data. Instead, through SSL, the goal is to learn representative features that are useful for the downstream task. Consequently, our goal in this paper is to devise data augmentation techniques and contrastive loss functions that can help extract meaningful features from the unlabeled data---which in our case correspond to noisy speech.}

In place of the unavailable clean speech references, both SSL procedures repurpose the target speaker's noisy speech as the new training target.
The first method we investigate is \textit{pseudo speech enhancement} (PseudoSE), referred to in other literature as \textit{noisy target training}.
We mix further noise onto the already-noisy observations and train the PSE model to remove the newly injected noise.
This self-supervised task is mismatched from true speech enhancement since the input signals are doubly degraded.
Therefore, the upper bound of PseudoSE is determined by how noisy the in-the-wild data originally is.
Nevertheless, specialist models trained using PseudoSE have an advantage over generalists because their training data is \textit{in-domain}, i.e., the noisy speech signals are still speaker-specific as opposed to speaker-agnostic.

The second method we investigate is \textit{contrastive mixtures} (CM).
Here, we reorganize the doubly-degraded input signals into pairs.
We prepare positive pairs---which share the same noisy speech source but have different injection noises.
Conversely, the negative pairs have identical injection noises but differing noisy speech sources.
Once the paired inputs are pseudo-denoised, the PSE model must learn to maximize the similarity of positive pair outputs and minimize the similarity of negative pair outputs.
CM improves upon PseudoSE thanks to the contrastive learning terms. 
Negative pairs, in particular, provide an additional learning opportunity.
Note that speech enhancement can be seen as a subset of source separation, wherein the two sources are known to be speech and noise.
Any additional utilization of the source separation nature of the problem will be a plus.
Hence, with CM, when preparing negative pairs, we see the noise source as the shared primary source while the speech sources are considered interfering sources that disagree.
Another view is that the negative pairs' disagreement objective acts as a regularization of PseudoSE.

Additionally, we show that both SSL frameworks can benefit from a \textit{data purification} (DP) process.
As its name suggests, DP makes the unlabeled noisy speech signals more useful by identifying the parts that contain purer speech than others.
In that regard, DP can be seen as an ``active learning'' method \cite{Gal2017active_learning} where the goal is to focus on more important samples from a large unlabeled dataset.
Regarding speech enhancement, we utilize DP by training a neural network in advance to predict speech quality frame-by-frame.
Prior research has shown the feasibility of predicting the long-term signal-to-noise ratio (SNR) of a noisy speech signal \cite{DongX2018lvaica}.
Our quality predictor, which estimates segmental SNR, identifies the relatively cleaner segments within all the noisy unlabelled data, a process similar to auto-labeling.
We convert the quality predictor's estimates into weights which guide the personalized speech enhancement model's loss function to prioritize cleaner input frames \cite{SivaramanA2021interspeech}.
Then, the PseudoSE function becomes refined through data purification: the frames of audio that are doubly degraded become diminished through weighting.
Consequently, the PseudoSE function will resemble an ideal speaker-specific fully-supervised learning process where only a single noise injection remains.

In summary, this research study explores personalized speech enhancement (PSE) as either a zero-shot (ZSL) or few-shot (FSL) learning problem.
We investigate two self-supervised (SSL) methods for training PSE models, and we augment both methods using data purification (DP).
Through our experiments, we assess the efficiency of the training methods in terms of data and model complexity.
Notably, in the FSL context, data efficiency is achieved by developing models which utilize as little user-provided clean speech as possible.
To this end, our experiment results show that models pretrained using the proposed self-supervised methods see greater improvements using a smaller amount of clean speech as opposed to fully-supervised generalist pretrained models.
Our experiments also assess PSE models of four different sizes, reinforcing the idea that in-domain self-supervised training allows smaller models to achieve a more competitive speech enhancement performance.

\section{Related Works}

We treat PSE as a data-driven problem, where training using in-domain data implicitly informs the model about the target speaker and their environment.
This differs from other studies, which tackle model personalization using auxiliary metadata, e.g., speaker-identifying embedding vectors \cite{WangQ2018voicefilter, GiriR2021interspeech, EskimezSE2021personalized}.

As mentioned in Sec. \ref{sec:intro}, we posit that PSE models benefit over SE models for optimized performance but also for lossless model compression.
Prior research has empirically shown that model specialization does lead to performance gains \cite{KolbaekM2017ieeeacmaslp, KimMJ2017icassp}. 
Intuitively, focusing on a small subset of the initially complex problem simplifies the neural network learning objective.
Therefore, with voice-controlled devices, we envision that a truly personalized speech enhancement model can optimally improve the experience of its primary user, forgoing other speakers and other unlikely acoustic scenarios.

Although model compression is an active area in deep learning research, many standardized methods, such as quantization or pruning \cite{HanS2016iclr}, do not consider the context of the model after deployment. Decreasing the total number of model parameters without reformulating the model objective is an option, but this may result in discernible performance trade-offs \cite{HowardAG2017Mobilenets}. Particularly with regards to speech enhancement or speech separation, more recent research has focused on novel model compression methods, including bitwise operations \cite{KimMJ2018icassp, GuoL2018icassp, KimSW2019icassp, KimSW2020icassp} or group communication in intermediate neural network layers \cite{LuoY2021icassp}. These works successfully minimize the performance trade-off but miss the opportunity to exploit the model's deployment environment. With personalization, because the sub-problem is easier to solve, a compressed specialist model suffices to perform on par with a more complex generalist model. For example, using a model selection approach conditioned on speaker genders, a specialist using \num{512} hidden units produced speech signal improvement comparable to a generalist model using \num{1024} hidden units, yielding an effectively ``lossless'' \SI{50}{\percent} reduction in run-time computational complexity \cite{SivaramanA2020interspeech}.

Model selection is another approach for designing adaptive models.
Four recent studies investigated run-time model selection as another means for test-time adaptation \cite{SivaramanA2020interspeech, ZezarioRE2020zeroshotse, ChazanS2021mle, SivaramanA2021waspaa}.
To develop a speech enhancement network through model selection, one must first cluster a large training corpus into non-overlapping subsets. 
Next, separate specialist submodules must be trained, each optimized around one of those subsets.
Lastly, the network requires a classifier that assigns the unseen speaker's noisy utterances to the best-suited submodule at test-time. 
Sparse activation of these submodules helps to optimize performance and reduce run-time computational complexity \cite{SivaramanA2020interspeech}.
By its design, model selection is inherently a zero-shot solution because the enhancement network does not utilize any prior knowledge about the target user.
Rather, using a noise-robust classifier and choosing the best submodule proxies adaptation.
We note that if all the submodules are very active during test-time, then the memory footprint savings of model selection are limited.
Additionally, the network's spatial complexity linearly scales with the number of clusters, making model selection impractical for edge computing devices.

Another recent study explored knowledge distillation as a proxy for test-time personalization that can overcome all of the pitfalls mentioned above \cite{KimSW2021waspaa}.
This dual-network configuration works by first training a teacher model, which processes test-time noisy signals from the unknown target speaker to produce pseudo-clean speech targets.
A student model (with significantly fewer parameters than the teacher model) must learn using these pseudo targets.
Through this framework, the student model is adapted to the target speaker's characteristics and target environment.
We must recognize that the student model's performance is fundamentally upper-bounded by the teacher model's, favoring a larger and more powerful teacher.
If the teacher model is prohibitively large, it must be placed off-device, which entails online-offline parameter updating procedures.
Thus, in noting the limitations of model selection and knowledge distillation, we put forward resource-efficient methods for personalizing a model in this paper.

Other research in the last few years has explored SSL for general-purpose speech enhancement.
An early work employed zero-shot SSL in a student-teacher framework, showing a student network that implicitly learned to perform speech enhancement despite being trained to minimize automatic speech recognition error \cite{WatanabeS2017student}.
More recently, another work describes an SSL framework that uses two autoencoders, trained to reproduce either clean speech or noisy speech \cite{WangYC2020selfsupervised}.
The authors enforce a coupling of the two autoencoders' latent spaces through cycle-consistency loss functions.
At inference time, the autoencoder trained only using mixture signals has its decoder swapped out, thus achieving zero-shot speech enhancement.
These studies are limited to speaker-agnostic enhancement, and in particular, do not exploit self-supervised learning as a method for in-domain training.
\change{
In contrast, two recent studies investigated using noisy speech data as target signals specifically for in-domain speech enhancement training \cite{MaciejewskiM2021icassp, FujimuraT2021eusipco}.
The PseudoSE method of this paper is similar to what they propose, however, our study investigates the benefits of noisy training targets specifically with regards to single-speaker model personalization and model compression.
Additionally, our study is the first to bootstrap noisy target training with contrastive learning with regards to speech enhancement.
}

Recently, an SSL framework known as mixture invariant training (MixIT) \cite{WisdomS2020mixit} was proposed as an alternative to the fully-supervised permutation invariant training (PIT).
It is a procedure for developing source separation systems using only mixtures of mixtures (MoM), i.e., linear combinations of arbitrary audio signals.
Considering MixIT as a pretext task, it introduces systematic mismatch by design since the input MoMs have twice the number of expected sources at test-time.
A recent study used MixIT by successfully adapting models to a set of speakers through joint training over in-domain and out-of-domain data \cite{SivaramanA2022icassp}, however the model compression implications were unexplored.
In comparison to MixIT, the PseudoSE task may be viewed as a more speech enhancement-oriented version: while MixIT estimates every composite signal, PseudoSE learns explicitly from the combination of a target speaker's noisy utterance plus an injection noise.
Therefore, a PseudoSE model is able to target the \textit{pseudo} speech source and can omit reconstructing the injection noise.

% \begin{figure}[h]
%     \centering
%     \includegraphics[width=\linewidth]{figures/jstsp_equations.pdf}
%     \caption{Overview of pretraining and finetuning procedures.}
%     \label{fig:equations}
% \end{figure}
% \input{tables/tbl_methods}

\section{Baseline Fully-Supervised Method for Personalized Speech Enhancement}

% We envision personalized speech enhancement applied in zero-shot and few-shot machine learning contexts. For example, many smart devices prompt users to register their voices. In this setting, the personalization target may provide a small amount of clean speech, on the order of seconds. If these signals are available in good quality, we consider this as a \textit{few-shot} learning (FSL) problem. However, if user-provided speech signals are unavailable, adapting to the unknown test-time speaker is a \textit{zero-shot} learning (ZSL) problem. For this paper, we limit the scope of personalization specifically regarding the test-time speaker and not the test-time environment. The extension of our methods towards environment adaptation is straightforward.

For our discussion, we assume a hypothetical set $\SSfull$ that encompasses all of the target speaker's clean utterances. \change{Given the privacy concerns and technical difficulties mentioned in Sec. \ref{sec:intro}, we assume that} this set is \change{inaccessible to the training algorithm}; therefore, it cannot be used for personalization. In the FSL context, the short recordings provided by the target speaker represent a small subset of their unavailable ground-truth clean speech, i.e., $\SStrain \in \SSfull$. The simplest approach for developing a personalized speech enhancement model would be to formulate a fully-supervised task over this subset. However, we theorize that the limited amount of data may result in suboptimal generalization performance and over-fitting. To remedy this issue, instead of randomly initializing the personalized model's parameters, one can first train a speaker-agnostic model and then finetune its parameters using $\SStrain$. Then, using transfer learning, we adapt a generalist model into a specialist model.

\subsection{Training a Generalist}
\label{sec:generalist}

Training a generalist requires a large set of many anonymous speakers $\SAfull$ as well as a large set of various non-stationary noises $\Nfull$. A training set of artificial mixture signals $\bm{x}$ can be made by selecting random utterances $\bm{s} \in \SAtrain$ and noises $\bm{n} \in  \Ntrain$ and summing the signals, i.e. $\bm{x} = \bm{s} + \bm{n}$. With each mixture, one may randomly scale $\bm{n}$ to be louder or quieter, thereby exposing the model to mixtures with varying signal-to-noise ratios (SNR). The generalist model can be described as a mapping function $f(\cdot)$ with parameters $\mathcal{W}_\text{SE}$ which is trained such that $f(\bm{x}; \mathcal{W}_\text{SE})= \bm{y} \approx \bm{s}$, where the estimate $\bm{y}$ approximates the training target $\bm{s}$. The generalist's loss function $\mathcal{L}_\text{SE}$ is equivalent to the discrepancy between estimates and targets: $\dist{\bm{y}}{\bm{s}}$. 
\begin{align}
    \mathcal{L}_\text{SE} &= \dist{\bm{y}}{\bm{s}} \\ \label{eq:loss_baseline}
    \displaystyle\mathcal{W_\text{SE}} &\leftarrow \argmin_{\mathcal{W}_\text{SE}} \mathcal{L}_\text{SE}
\end{align}
\change{There are many possible choices for the signal discrepancy function $\mathcal{E}$. The well-known signal-to-distortion ratio (SDR) metric \cite{VincentE2006ieeeaslp} is frequently used as a general-purpose loss function for fully-supervised monaural time-domain speech enhancement \cite{SaitoK2021noisyspeech}.
A larger SDR correlates to improved speech quality, so when used as a neural network loss function, we minimize the negative of SDR.
For a source signal $\bm{v}$ and estimate signal $\hat{\bm{v}}$, negative SDR loss is defined as follows:
\begin{align}
\mathcal{E}_{\operatorname{SDR}} (\hat{\bm{v}} \;\|\; \bm{v}) 
=
-10 \log_{10}\left[ \frac{\sum_t (v_t)^{2} }{\sum_t (v_t - \hat{v}_t )^{2} } \right].
\label{eq:sdr}
\end{align}}

For generalists, what matters most is their generalization power. Although synthetic mixtures for fully-supervised training are straightforward to construct, models with low architectural complexity may not learn much from the data. That is, a smaller model may fail to enhance certain speakers' voices or remove particular noises---even if the training corpora for speech and noise signals were very large. In contrast, a bigger model may generalize very well, but using it in a resource-constrained device could be burdensome.

\begin{figure}
    \centering
    \def\arraystretch{2}
    \setlength\tabcolsep{2pt}
    \begin{tabular}{rcl}
    & \includegraphics[align=c]{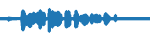}
    & $\bs \in \SAtrain$ \\
    {\small Noise Injection}
    & \includegraphics[align=c]{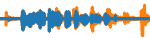}
    & $\bx=\bs+\bn; \hspace{1mm} \bn \in \Ntrain$ \\
    {\small Enhancement}
    & \includegraphics[align=c]{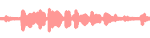}
    & $\bm{y}=f\left(\bm{x} ; \mathcal{W}_{\text{SE}}\right)$ \\
    \end{tabular}
    \caption{Multi-speaker (fully-supervised) speech enhancement setup. The training target is clean speech $\bm{s}$ and the model parameters $\mathcal{W}_\text{SE}$ are iteratively updated to minimize the loss function $\mathcal{L}_\text{SE}$. In the FSL context, we can finetune the model by sampling $\bs$ from the small speaker-specific dataset $\SStrain$.}
    \label{fig:eq_baseline}
\end{figure}

\subsection{Personalization via Transfer Learning}

The speaker-agnostic speech enhancement model may then be finetuned around the particular test-time speaker using transfer learning. Transfer learning is a straightforward fully-supervised approach to personalization, which handles the gap between the large multi-speaker dataset $\SAfull$ and the small target speaker-provided clean dataset $\SStrain$. To do this, we create speaker-specific artificial mixture signals $\bm{x}$ composed stochastically by sampling from the limited subset $\bm{s} \in \SStrain$ and the training noises $\bm{n} \in \Ntrain$. The parameters $\mathcal{W}_\text{SE}$ are once again iteratively updated in order to minimize the distance between estimate signals $\bm{y}$ and target signals $\bm{s}$. The finetuning loss function is equivalent to Eq. \eqref{eq:loss_baseline}, but during finetuning, the model receives exposure to utterances from the target speaker.

The success of transfer learning as a personalization method depends on how effective the pretraining and finetuning steps are. For example, a large model highly generalized thanks to pretraining might barely adjust its parameters during finetuning. On the other hand, smaller models with weaker generalization capabilities may see a more significant performance boost through finetuning. Ultimately, the success of finetuning is primarily tied to the quality and quantity of the finetuning dataset $\SStrain$. Suppose the number of signals within $\SStrain$ is too few; in that case, finetuning may fail to improve performance even though $\SStrain$ consists of the target speaker's vocal characteristics. Also, because the FSL context only applies when the target speaker manually provides their clean speech, transfer learning is not viable without $\SStrain$.

Fig. \ref{fig:eq_baseline} shows a visualization of the baseline pretraining process. The same signal transformations occur during transfer learning, when adapting the generalist model into a specialist model. If the target speaker does not provide $\SStrain$, the generalist model remains unadapted and therefore non-personalized.

\section{Proposed Self-Supervised Methods for Personalized Speech Enhancement}

Here we describe our proposed self-supervised learning (SSL) methods, designed to improve the performance of the personalized speech enhancement models in either FSL or ZSL contexts. Through SSL, we aim at pretraining an SE model that can surpass the performance of the baseline generalist. This pretraining can suffice as a personalized solution (i.e., ZSL). Or, we can further finetune the self-supervised model by using the small amount of target speech signals if they are available (i.e., FSL). 

Our utilization of SSL stems from the assumption that \textit{noisy} utterances from the target speaker $\tilde{\bm{s}} \in \SSnpretrain$ are much more available than clean ones, i.e., $|\SSnpretrain| \gg |\SStrain|$. Our proposed pretraining methods aim to exploit these noisy observations as much as possible to learn the specificity of the test-time speaker. As is the case with SSL methods, the model parameters will be initialized via a pretext task, which is a made-up task that does not reflect a true speech enhancement function.

We assert, for example, that smart devices are likely to accrue many noisy recordings from the test-time speaker over time and with usage, i.e., $|\SSnpretrain|\gg|\SStrain|$.
Although we want to exploit these in-the-wild recordings $|\SSnpretrain|$, we do not know whether the observations are clean or noisy, i.e., the data is unlabeled.
Therefore, we have to assume that $|\SSnpretrain|$ holds contaminated versions of some unobserved target clean speech signal $|\SSpretrain|$. We refer to this unobserved contamination process as \textit{premixture}.
If we consider a hypothetical set of premixture noises $\bm{m} \in \Mtrain$, then we can form a basic framework for premixture, i.e., $\tilde{\bm{s}} = \bm{s} + \bm{m}$. 
Because the true speech and noise signals which compose $\tilde{\bm{s}}$ are unknown, the premixture observations are unsuitable for conventional fully-supervised speech enhancement tasks nor for finetuning-based personalization.

\begin{figure*}[t]
\includegraphics[width=.95\textwidth]{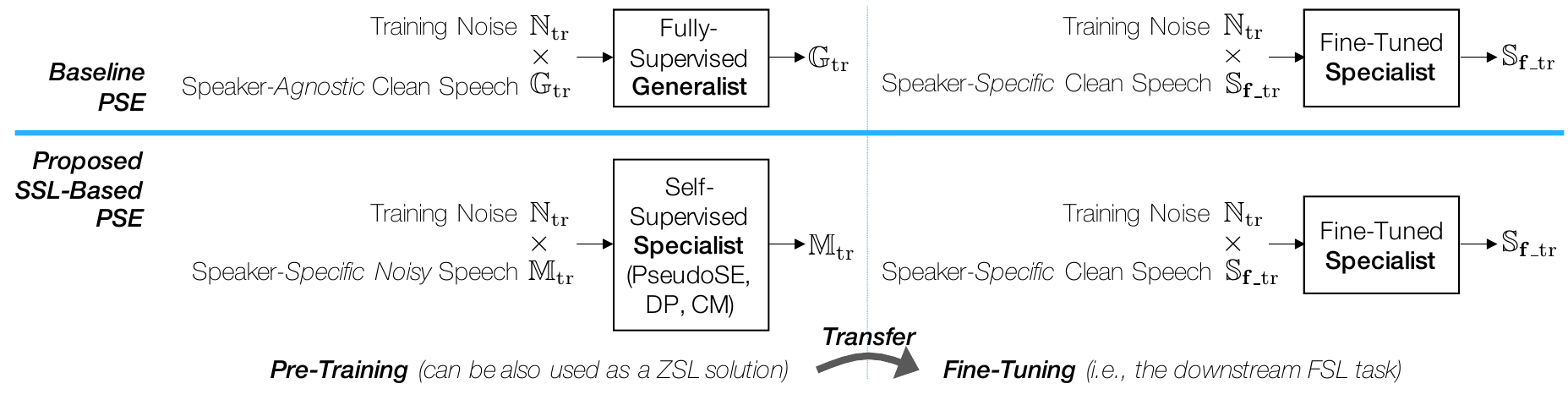}
\caption{\change{An overview of the baseline and proposed personalization methods. With the baseline, the SE model is first pretrained using speaker-agnostic dataset as a generalist and then finetuned using clean speech signals of the test user. This method relies entirely on the finetuning process for personalization. On the other hand, the proposed methods provide various SSL options to pretrain the model using noisy, but speaker-specific speech, which serve a better initialization point for the subsequent finetuning process, leading to better SE performance. The pretrained models can also conduct a certain level of SE as a ZSL model, while the FSL-based finetuning tends to improve the pretrained model.}}
\label{fig:overview}
\end{figure*}

\change{Fig. \ref{fig:overview} summarizes the training procedure of the baseline generalist-based pretraining, comparing it to our proposed SSL-based pretraining. Both approaches to personalization are based on transfer learning. Finetuning via FSL improves the baseline SE performance, exposing the generalist to the target speaker. However, the proposed SSL methods already achieve a certain level of personalization by using noisy speech signals of the target speaker, leading to a better ZSL solution than the generalist.}

% \begin{table}[h!]
% \newcommand\cw{3.2in} % equation column
% \centering
% \raggedright{
% \setlength{\tabcolsep}{2pt}
% \def\arraystretch{1.5}
% {\normalsize
% \begin{tabular}{m{1ex}m{4pt}m{\cw}m{1in}}
% \multicolumn{4}{l}{
% \textbf{Premixture} (Simulation of sampling from $\SSnpretrain$)
% } \\
% && \inlineequation[eq:premixture]{
% \tilde{\bm{s}} = \bm{s} + \bm{m}; \quad \bm{s}\in \SSpretrain, \hspace{2mm} \bm{m} \in \Mtrain}
% \end{tabular}
% \begin{tabular}{m{1ex}m{4pt}m{\cw}m{1in}}
% \multicolumn{4}{l}{
% \textbf{Specialist pretraining} (Self-Supervised / ZSL)
% } \\
% \multirow{3}{*}{\rotatebox[origin=c]{90}{\textcolor{tab_blue}{PseudoSE}}}
% & \multirow{3}{*}{\color{tab_blue}\vline height 52pt depth 0pt width 2pt}
% & $\tilde{\bm{x}} = \tilde{\bm{s}} + \bm{n}; \quad \tilde{\bm{s}}\in\tilde{\mathbb{S}}_\text{\textbf{p}-tr},\hspace{2mm}\bm{n}\in\Ntrain$
% \\
% && $\tilde{\bm{y}} = f_\text{PSE}(\tilde{\bm{x}}; \mathcal{W}_\text{PseudoSE})$
% \\
% && \inlineequation[eq:obj_pseudose]{
% \mathcal{L}_\text{PSE} = \dist{\tilde{\bm{y}}}{\tilde{\bm{s}}}}
% \end{tabular}
% }}
% \end{table}

\begin{figure}
    \centering
    \def\arraystretch{2}
    \setlength\tabcolsep{2pt}
    \begin{tabular}{rcl}
    & \includegraphics[align=c]{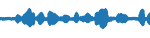}
    & $\bs \in \SSpretrain$ \\
    {\small Premixture}
    & \includegraphics[align=c]{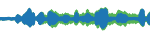}
    & $\tilde{\bm{s}} = \bs + \bm{m}; \hspace{1mm} \bm{m}\in\Mtrain$ \\
    {\small Noise Injection}
    & \includegraphics[align=c]{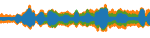}
    & $\tilde{\bm{x}} = \tilde{\bm{s}} + \bm{n}; \hspace{1mm} \bn \in \Ntrain$ \\
    {\small Enhancement}
    & \includegraphics[align=c]{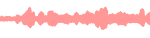}
    & $\tilde{\bm{y}} = f\left(\bm{x} ; \mathcal{W}_{\text{PseudoSE}}\right)$ \\
    \end{tabular}
    \caption{Single-speaker (self-supervised) pseudo speech enhancement setup. The training target is pseudo-clean speech $\tilde{\bm{s}}$, therefore the model parameters $\mathcal{W}_{\text{PseudoSE}}$ are iteratively updated to minimize the loss function $\mathcal{L}_{\text{PseudoSE}}$. We simulate the process of sampling from the in-the-wild recordings, $\tilde{\bs} \in \tilde{\mathbb{S}}_{\textbf{p}\text{-tr}}$, using the premixture data transformation.}
    \label{fig:eq_pseudose}
\end{figure}

\subsection{Training a Specialist through Pseudo Speech Enhancement}\label{sec:PseudoSE}

Depending on the user's test-time acoustic conditions, it is likely that the premixture noise component $\bm{m}$ has a loudness that varies over time.
Then it follows that, at certain times, this premixture noise may be quiet enough such that the test-time speaker's voice $\bm{s}$ is the dominant signal.
In these cases where there is a favorable premixture with a high signal-to-noise ratio (SNR), the noisy speech utterances $\tilde{\bm{s}}$ could be used as \textit{pseudo} speech references.
We can then formulate a pretraining process which we call \textit{pseudo speech enhancement} (PseudoSE).
This method operates using ``doubly-degraded'' artificial mixture signals.
We construct the model inputs by sampling the abundant premixture set $\tilde{\bm{s}} \in \SSnpretrain$ and injecting the additional training noises $\bm{n} \in \Ntrain$, i.e., $\tilde{\bm{x}} = \tilde{\bm{s}} + \bm{n}$.
This is a double-degradation process as $\tilde{\bm{s}}$ has been already contaminated by $\tilde{\bm{m}}$. 

Consequently, the self-supervised model is a mapping function $f$ with parameters $\mathcal{W}_\text{PseudoSE}$ that is trained to remove the injection noise and recover the pseudo speech target, i.e., $f(\bm{\tilde{x}}; \mathcal{W}_\text{PseudoSE}) = \tilde{\bm{y}} \approx \tilde{\bm{s}}$.
Note that this self-supervised objective is not equivalent to the fully-supervised objective due to the difference in training target.
$f$ is only trained to recover the premixture utterance $\tilde{\bm{s}}$, therefore it is not a true speech enhancement function, i.e., $\mathcal{W}_\text{PseudoSE} \neq \mathcal{W}_\text{SE}$. 
\begin{align}
    \label{eq:loss_pseudose}
    \mathcal{L}_\text{PseudoSE} &= \dist{\tilde{\bm{y}}}{\tilde{\bm{s}}} \\
    \displaystyle\mathcal{W_\text{PseudoSE}} &\leftarrow \argmin_{\mathcal{W}_\text{PseudoSE}} \mathcal{L}_\text{PseudoSE}
\end{align}

% With all of the discussed pretraining methods, we use the scale-invariant signal-to-noise ratio (SI-SNR) as the discrepancy function $\mathcal{E}$. Given estimate signals $\by$ and target signals $\bs$, the machine learning models update their parameters with each iteration in order to minimize the MSE terms.
% \begin{align}
% \mathcal{E}\distx{\by}{\bs} = \operatorname{SI-SNR}(\by, \bs) = \sum_i\left(s_i - y_i\right)^{2},
% \tag{9} \label{eq:mse}
% \end{align}
% where $i$ denotes the index of time-domain samples.

Fig. \ref{fig:eq_pseudose} shows a visualization of the PseudoSE pretraining process. After the model parameters $\mathcal{W}_\text{PseudoSE}$ are learned, we may apply finetuning using known clean speech from the scarce set $\SStrain$. In this FSL personalization context, the training targets are genuine clean speech utterances $\bs\in\SStrain$. Therefore, the parameters from the pseudo enhancement function $\mathcal{W}_\text{PseudoSE}$ are iteratively updated in order to fit a real speech enhancement function. Once again, the finetuning loss function is equivalent to Eq. \eqref{eq:loss_baseline} using the speaker-specific mixtures.

% \begin{align}
%     \begin{split}
%     \label{eq:pse_pretrain}
%     \text{pretraining: }
%     & \tilde{\bm{s}} = \bm{s} + \bm{m}; \quad \bm{s}\in\mathbb{S}_\text{p-tr},\hspace{2mm}\bm{m}\in\mathbb{M}_\text{tr} \\
%     & \tilde{\bm{x}} = \tilde{\bm{s}} + \bm{n}; \quad \bm{n} \in \mathbb{N_\text{tr}} \\
%     & \argmin_{\mathcal{W}_f} \mathcal{E}( \tilde{\bm{y}} = f(\tilde{\bm{x}}; \mathcal{W}_f)||\tilde{\bm{s}})
%     \end{split}
% \end{align}

% 
% 12/21 03:00 Aswin
% 
There are trade-offs to note when using self-supervised learning.
On one hand, the success of PseudoSE pretraining is bounded by the noisiness of $\tilde{\bm{s}}$, the impure training targets.
But on the other hand, this pretraining scheme uses data derived only from the target speaker, thereby bypassing the need for generalization. 
Unlike the baseline method, which recasts a generalist as a specialist, PseudoSE pretraining directly develops a specialist model.
However, the PseudoSE model could under perform when compared to a hypothetical fully-supervised model exposed to ample clean speech from the target speaker.
If finetuning is not possible, the PseudoSE model could serve as a zero-shot solution on its own.
But if finetuning is possible, we claim that PseudoSE serves as a more optimal pretraining scheme as opposed to the baseline speaker-agnostic SE.

\subsection{Training a Specialist through Contrastive Mixtures}
\label{sec:contrastive}

\begin{figure}[t]
\centering
\subfloat[Positive Pairs]{\includegraphics[width=.47\columnwidth]{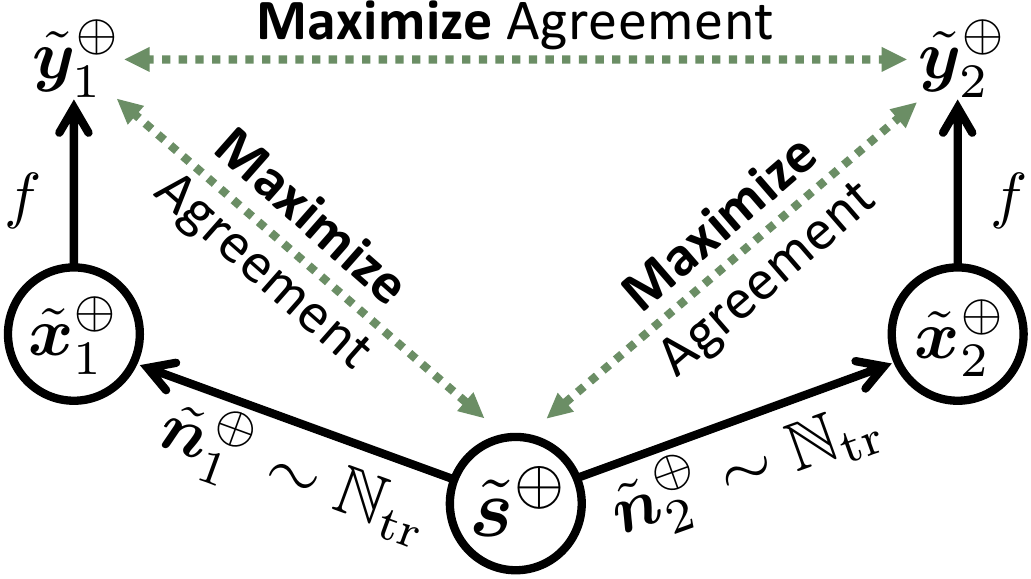}} \hfill
\subfloat[Negative Pairs]{\includegraphics[width=.47\columnwidth]{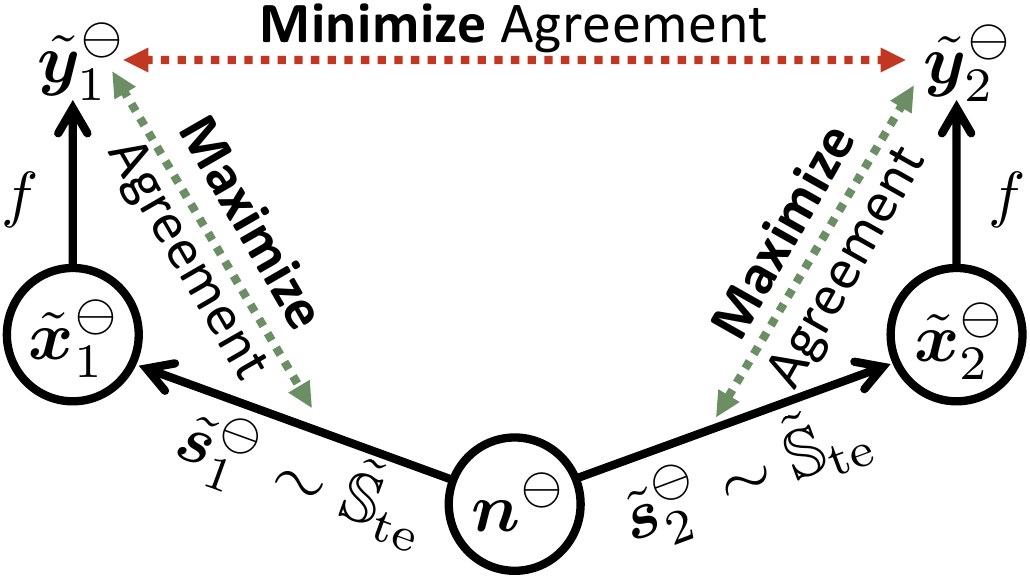}}
\caption{The proposed framework for contrastive mixtures. Solid lines indicate signal path while dashed lines show loss terms.}
\label{fig:cm_framework}
\end{figure}

% \begin{align}
%     \begin{split}
%     \label{eq:cm_pretrain}
%     \text{pretraining: }
%     & \tilde{\bm{s}} = \bm{s} + \bm{m}; \quad \bm{s}\in\mathbb{S}_\text{p-tr},\hspace{2mm}\bm{m}\in\mathbb{M}_\text{tr} \\
%     & \tilde{\bm{x}}_1 = \tilde{\bm{s}} + \bm{n}_1; \quad \bm{n}_1 \in \mathbb{N_\text{tr}} \\
%     & \tilde{\bm{x}}_2 = \tilde{\bm{s}} + \bm{n}_2; \quad \bm{n}_2 \in \mathbb{N_\text{tr}} \\
%     & \bm{y}_1 = f(\tilde{\bm{x}}_1 ; \mathcal{W}_f) \\
%     & \bm{y}_2 = f(\tilde{\bm{x}}_2 ; \mathcal{W}_f) \\
%     & \argmin_{\mathcal{W}_f}
%     \begin{aligned}[t]
%       & \left\big[ \mathcal{E}( \tilde{\bm{y}}_1||\tilde{\bm{s}})
%       + \mathcal{E}( \tilde{\bm{y}}_2||\tilde{\bm{s}}) \right. \\
%       & \left. \quad + \lambda \cdot \mathcal{E}( \tilde{\bm{y}}_1||\tilde{\bm{y}}_2) \right\big]
%     \end{aligned}
%     \end{split} \\[1ex] \begin{split}
%     \label{eq:cm_finetune}
%     \text{finetuning: }
%     & \mathcal{W}_\text{SE}=\mathcal{W}_f \\
%     & \bm{x} = \bm{s} + \bm{n}; \quad \bm{s}\in\mathbb{S}_\text{f-tr},\hspace{2mm}\bm{n}\in\mathbb{N}_\text{tr} \\
%     & \argmin_{\mathcal{W}_\text{SE}} \mathcal{E}(\bm{y}=g(\bm{x}; \mathcal{W}_\text{SE})||\bm{s})
%     \end{split}
% \end{align}

We hypothesize that the quality of the pretraining procedure greatly impacts how the downstream denoising model will personalize. Even if the premixed noisy speech set $\SSnpretrain$ and the deformation noise set $\Ntrain$ are large, the quality of the features learned through PseudoSE are bounded by how noisy $\SSnpretrain$ really is. Our proposed \textit{contrastive mixtures} (CM) pretraining procedure addresses this by employing a pairwise contrastive learning mechanism. In the CM framework, the denoising model $f(\cdot)$ pretrains over \textit{pairs} of mixtures ($\tilde{\bm{x}}_1$, $\tilde{\bm{x}}_2$) and outputs pseudo-cleaned estimates ($\tilde{\bm{y}}_1$, $\tilde{\bm{y}}_2$). We create two kinds of mixture pairs, \textit{positive} and \textit{negative}, which are illustrated in Figure \ref{fig:cm_framework}.% with example spectrograms shown in Figures \ref{fig:positivepair_spec} and \ref{fig:negativepair_spec}.

In a positive pair, both input examples ($\tilde{\bm{x}}_1^\oplus$, $\tilde{\bm{x}}_2^\oplus$) share the same premixture source $\tilde{\bm{s}}^\oplus$, but are differently deformed; that is, the mixing process makes the input pair dissimilar. Therefore, in addition to maximizing the similarities between estimates and source ($\tilde{\bm{y}}_1^\oplus$ to $\tilde{\bm{s}}^\oplus$ and $\tilde{\bm{y}}_2^\oplus$ to $\tilde{\bm{s}}^\oplus$), the model $f_\text{CM}(\cdot)$ must also satisfy the contrastive objective based on the fact that $\tilde{\bm{y}}_1^\oplus$ and $\tilde{\bm{y}}_2^\oplus$ stemmed from the same pseudo source. We express these objectives as a positive pair loss function $\mathcal{L}_p$ in the following form:
\begin{equation}
    \mathcal{L}_p =
    \mathcal{E}(\tilde{\bm{s}}^\oplus||\tilde{\bm{y}}_1^\oplus) + \mathcal{E}(\tilde{\bm{s}}^\oplus||\tilde{\bm{y}}_2^\oplus) +
    \lambda_p \big[\mathcal{E}(\tilde{\bm{y}}_1^\oplus||\tilde{\bm{y}}_2^\oplus)\big],
\label{eq:loss_positive}
\end{equation}
where $\lambda_p$ scales the contribution of the contrastive loss term. 

In a negative pair, each mixture is made from a \textit{different} pseudo source ($\tilde{\bm{s}}_1^\ominus\neq \tilde{\bm{s}}_2^\ominus$), but with a shared deformation, i.e., $\tilde{\bm{x}}_1^\ominus=\tilde{\bm{s}}_1^\ominus+{\bm{n}}^\ominus$ and $\tilde{\bm{x}}_2^\ominus=\tilde{\bm{s}}_2^\ominus+{\bm{n}}^\ominus$; in other words, the negative pair mixing process makes the originally different inputs more similar to one another. Accordingly, in addition to the source-wise denoising objectives, \change{the dissimilarity between the estimates $\tilde{\bm{y}}_1^\ominus$ and $\tilde{\bm{y}}_2^\ominus$ must be taken into consideration. We express these objectives as a negative pair loss function $\mathcal{L}_n$ in the following form:
\begin{equation}
\begin{split}
    \mathcal{L}_n &= \mathcal{E}(\tilde{\bm{s}}_1^\ominus||\tilde{\bm{y}}_1^\ominus) + \mathcal{E}(\tilde{\bm{s}}_2^\ominus||\tilde{\bm{y}}_2^\ominus)\\
    &+ \lambda_n \big[ \operatorname{max} \big( \mathcal{E}(\tilde{\bm{s}}_1^\ominus||\tilde{\bm{s}}_2^\ominus), \mathcal{E}(\tilde{\bm{y}}_1^\ominus||\tilde{\bm{y}}_2^\ominus) \big)\big],
\label{eq:loss_negative}
\end{split}
\end{equation}
where $\lambda_n$ controls the contribution of the contrastive loss term. Note that the $\operatorname{max}$ function sets up the bound for the disagreement term $\mathcal{E}(\tilde{\bm{y}}_1^\ominus||\tilde{\bm{y}}_2^\ominus)$ comparing it with the ``desired" disagreement level of the target pseudo sources $\mathcal{E}(\tilde{\bm{s}}_1^\ominus||\tilde{\bm{s}}_2^\ominus)$, rather than enforcing an unbounded disagreement.}

Both $\mathcal{L}_p$ and $\mathcal{L}_n$ consist of two terms: the source-to-estimate errors and the estimate-to-estimate errors. The former term characterizes the main speech enhancement loss, while the latter term provides the proposed contrastive regularization. The model ultimately minimizes the sum of these two losses,
\begin{align}
\mathcal{L}_\text{CM} &= \sum_{t=1}^T\mathcal{L}_p(t) + \sum_{t=1}^T\mathcal{L}_n(t) \label{eq:loss_contrastive}\\
    \displaystyle\mathcal{W_\text{CM}} &\leftarrow \argmin_{\mathcal{W}_\text{CM}} \mathcal{L}_\text{CM},
\end{align}
where $T$ is the number of positive or negative pairs within the batch and $\mathcal{L}_p(t)$ and $\mathcal{L}_n(t)$ denote the loss for the $t$-th pair.
If the regularizing contrastive terms are omitted, i.e., by setting $\lambda_p = 0$ and $\lambda_n = 0$, it can be shown that $\mathcal{L}_\text{CM}$ reduces to Eq. \eqref{eq:loss_pseudose}.
Four our experiments, we set $T$ to be half of the batch size.
To find optimal choices for $\lambda_p$ and $\lambda_n$, we run an ablation study as described in Sec. \ref{sec:ablation}.

% We performed an ablation study to investigate the contribution of the contrastive loss terms for both positive and negative pairs. As seen in Fig. \ref{fig:ablation_study}, we found that having both $\lambda_p = 1$ and $\lambda_n = 1$ minimized training loss, consequently maximizing test-time enhancement. Note that the impact of negative pairs is minimal in the small models while they start to show improvements in medium-sized models.

Our proposed CM approach differs from the SimCLR model \cite{ChenT2020simclr} in multiple regards: (a) it uses a more sophisticated noise injection for data augmentation to mimic the real-world noisy speech mixture generation process, i.e. by using non-stationary noise sources; (b) the introduction of the negative pairs more precisely reflects the source separation concept underlying our SE problem and yields a more discriminative feature than a positive pair only; and, (c) having the traditional SE loss term prevents trivial solutions to the contrastive loss-only case---estimating very similar $\tilde{\bm{y}}_1^\ominus$ and $\tilde{\bm{y}}_2^\ominus$ that do not recover the pseudo sources. 

\begin{figure}
    \centering
    \def\arraystretch{2}
    \setlength\tabcolsep{2pt}
    \begin{tabular}{rcl}
    & \includegraphics[align=c]{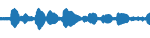}
    & $\bs \in \SSpretrain$ \\
    {\small Premixture}
    & \includegraphics[align=c]{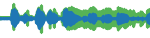}
    & $\tilde{\bm{s}}^\oplus = \bs + \bm{m}; \hspace{1mm} \bm{m}\in\Mtrain$ \\
    {\small Noise Injection}
    & \includegraphics[align=c]{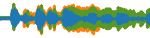}
    & $\tilde{\bm{x}}_1^\oplus = \tilde{\bm{s}}^\oplus + \bm{n}_1^\oplus; \hspace{1mm} \bm{n}_1^\oplus \in \Ntrain$ \\
    {\small Noise Injection}
    & \includegraphics[align=c]{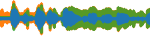}
    & $\tilde{\bm{x}}_2^\oplus = \tilde{\bm{s}}^\oplus + \bm{n}_2^\oplus; \hspace{1mm} \bm{n}_2^\oplus \in \Ntrain$ \\
    {\small Enhancement}
    & \includegraphics[align=c]{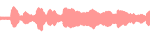}
    & $\tilde{\bm{y}}_1^\oplus = f\left(\tilde{\bm{x}}_1^\oplus ; \mathcal{W}_{\text{CM}}\right)$ \\
    {\small Enhancement}
    & \includegraphics[align=c]{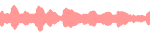}
    & $\tilde{\bm{y}}_2^\oplus = f\left(\tilde{\bm{x}}_2^\oplus ; \mathcal{W}_{\text{CM}}\right)$ \\[2ex]
    & \includegraphics[align=c]{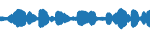}
    & $\bs_1 \in \SSpretrain$ \\
    {\small Premixture}
    & \includegraphics[align=c]{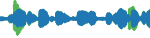}
    & $\tilde{\bm{s}}^\ominus_1 = \bs_1 + \bm{m}_1; \hspace{1mm} \bm{m}\in\Mtrain$ \\
    & \includegraphics[align=c]{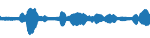}
    & $\bs_2 \in \SSpretrain$ \\
    {\small Premixture}
    & \includegraphics[align=c]{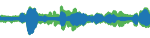}
    & $\tilde{\bm{s}}^\ominus_2 = \bs_2 + \bm{m}_2; \hspace{1mm} \bm{m}\in\Mtrain$ \\
    {\small Noise Injection}
    & \includegraphics[align=c]{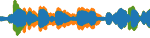}
    & $\tilde{\bm{x}}_1^\ominus = \tilde{\bm{s}}_1^\ominus + \bm{n}^\ominus; \hspace{1mm} \bm{n}_1^\ominus \in \Ntrain$ \\
    {\small Noise Injection}
    & \includegraphics[align=c]{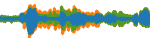}
    & $\tilde{\bm{x}}_2^\ominus = \tilde{\bm{s}}_2^\ominus + \bm{n}^\ominus$ \\
    {\small Enhancement}
    & \includegraphics[align=c]{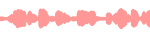}
    & $\tilde{\bm{y}}_1^\ominus = f\left(\tilde{\bm{x}}_1^\ominus ; \mathcal{W}_{\text{CM}}\right)$ \\
    {\small Enhancement}
    & \includegraphics[align=c]{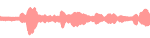}
    & $\tilde{\bm{y}}_2^\ominus = f\left(\tilde{\bm{x}}_2^\ominus ; \mathcal{W}_{\text{CM}}\right)$ \\
    \end{tabular}
    \caption{Single-speaker (self-supervised) contrastive mixtures setup. With positive pairs, there is a single training target, pseudo source $\tilde{\bm{s}}^\oplus$. With negative pairs, there are two different training targets, pseudo sources $\tilde{\bs}_1^\ominus$ and $\tilde{\bs}_2^\ominus$. The model parameters $\mathcal{W}_\text{CM}$ are iteratively updated to minimize the loss function $\mathcal{L}_\text{CM}$.}
    \label{fig:eq_cm_positive}
\end{figure}

% \begin{figure}
%     \centering
%     \includegraphics{figures/fig_eq_conmix_neg.pdf}
%     \caption{Single-speaker (self-supervised) contrastive pseudo speech enhancement setup.  The model parameters $\mathcal{W}_\text{CM}$ are iteratively updated to minimize the loss function $\mathcal{L}_n$.}
%     \label{fig:eq_cm_negative}
% \end{figure}

\subsection{Data Purification}

When it comes to fully-supervised pretraining, we know that the target signals are clean because they originate from the large labeled dataset $\SAtrain$. However, the target signals' cleanliness is ambiguous in the case of self-supervised pretraining, which utilizes $\SSnpretrain$ as the pseudo source. Based on our formulation of the premixture process in Fig. \ref{fig:eq_pseudose}, two factors determine whether the pseudo sources $\tilde{\bs}$ are too degraded to be usable. These are: the sparsity of premixture noise $\bm{m}$, as well as the segmental SNR between $\bm{s}$ and $\bm{m}$. For example, if $\bm{m}$ is sufficiently sparse, portions of $\tilde{\bs}$ may contain near-clean speech.
Considering all the available noisy utterances $\tilde{\bs} \in \SSnpretrain$, we hypothesize that utterances with a higher SNR may serve as more useful target signals than other noisier utterances, even if none of them are completely clean.
The proposed self-supervised pretraining methods can benefit from knowing where the cleaner frames within $\tilde{\bs}$ may be.

For that reason, we put forward a \textit{data purification} (DP) pipeline. In essence, we modify the discrepancy function $\mathcal{E}$ to incorporate a weighting vector $\bp$. To generate this DP weighting vector, we first train a separate neural network that estimates the frame-by-frame SNR of the premixtures. The quality estimator network $h$ is a regressive model trained over a diverse set of training speakers and noises (i.e., $\SAtrain$ and $\Ntrain$). It outputs a vector of segmental SNRs, $\hat{\bm{\alpha}}$. Hence, the network $h$ works as a general-purpose speech quality estimator, that has no prior knowledge of the test-time speaker or the test-time noisy environment. Given an estimate signal $\hat{\bm{v}}$ and a target signal ${\bm{v}}$ both of length $L$, their residual is $\bm{r} = {\bm{v}} - \hat{\bm{v}}$, and the frame-by-frame/segmental SNR (SegSNR) is defined as:
\begin{equation}
\operatorname{SegSNR}_j(\bm{v}, \hat{\bm{v}}) =
10 \log_{10}
\left[ \frac{
    \sum_{i=Hj}^{Hj+N-1} \left({w}_\change{i-Hj}{v}_{i}\right)^{2}
}{
    \sum_{i=Hj}^{Hj+N-1} \left({w}_\change{i-Hj}{r}_{i}\right)^{2}
} \right] ,
\label{eq:segmental_snr}
\end{equation}
where $N$ is the frame size, $H$ is the hop size, $j$ is a zero-based frame index (i.e. $0 \leq j \leq \ceil{\frac{L}{H}}-1$), and vector $\bm{w}$ comes from the Hann window function of length $N$. We then formulate the training process of the SNR Predictor network as follows:
\begin{align}
\begin{split}
     \bm{x}&=\bm{s}+\bm{n}; \quad \bm{s}\in\SAtrain, \hspace{1mm} \bm{n}\in\Ntrain \\
     \bm{\alpha}&=\operatorname{SegSNR}(\bm{s},\bm{x}) \\
     \hat{\bm{\alpha}}& = h(\bm{x}; \mathcal{W}_h) \\
     \mathcal{W}_h &\leftarrow \argmin_{\mathcal{W}_h} \hspace{1mm}
    \operatorname{MSE}(\hat{\bm{\alpha}}, \bm{\alpha}) , 
    % & \mathcal{W}_h \leftarrow \argmin_{\mathcal{W}_h} \hspace{1mm}
    % \operatorname{MSE} \Big(h(\bm{x}; \mathcal{W}_h), \hspace{1mm} \bm{\alpha}\Big) , 
\end{split}
\end{align}
Note that the SNR predictor inputs are of length $L$, but its outputs are of length $\ceil{\frac{L}{H}}$; in other words, $\bx$'s length is measured in samples but $\hat{\bm{\alpha}}$'s length is measured in frames.

\begin{figure}[t!]
    \centering
    \includegraphics[width=\linewidth]{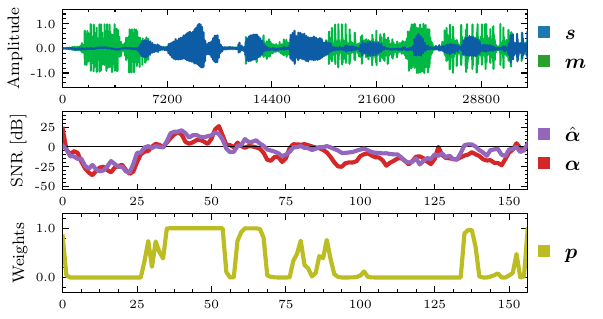}
    \caption{Illustration of the SNR predictor inputs and outputs. The first subplot features an example premixture/pseudo source $\tilde{\bm{s}}$. In the second subplot, the SNR predictor network $h$ estimates the frame-wise (i.e., segmental) SNR of the premixture. The training objective of $h$ is to minimize the loss between estimates $\hat{\bm{\alpha}}$ and targets $\bm{\alpha}$. The third subplot shows the frame-by-frame SNR estimates converted into weights using the logistic function, i.e. $\bm{p} = \sigma(h(\tilde{\bm{s}}))$.}
    \label{fig:snr_prediction}
\end{figure}

We can now apply a DP step to improve the reliability of the pseudo-target $\tilde{\bm{s}}$ during PseudoSE and CM pretraining. 
With each iteration of pretraining, the SNR predictor $h$ first analyzes the input premixtures to estimate frame-wise SNRs, $\hat{\bm{\alpha}} = h(\tilde{\bm{s}})$.
Next, we apply the logistic function $\sigma$ to the $\hat{\bm{\alpha}}$ logits in order to obtain frame-by-frame weights:
\begin{equation}
\bm{p}= \sigma(\hat{\bm{\alpha}}) = \frac{1}{1 + e^{-\hat{\bm{\alpha}}}}.
\end{equation}
\change{Lastly, we modify both PseudoSE and CM pretraining procedures to use $\mathcal{E}_\text{DP}$ which promotes speech-prominent frames in the loss function. To that end, we re-write Eq. \eqref{eq:segmental_snr} to incorporate the frame-by-frame weights $\bm{p}$. That is, the signal discrepancy is computed between windowed segments, which are then weighted by $\bm{p}$ and finally averaged across all frames. Because this is a neural network loss function to be minimized, we use the negative of weighted segmental SNR, which we denote as $\distDP$. 
% \begin{align}
% \begin{split}
% \mathcal{E}_\text{DP}\distx{\tilde{\bm{y}}}{\tilde{\bm{s}}}
% &= \operatorname{SegMSE}(\tilde{\bm{y}},\tilde{\bm{s}}; \bm{p}) \\
% &= \frac{1}{J} \sum_{j=0}^{J\!-\!1} p_{j}
% \left[ \frac{1}{N}  \sum_{i=Hj}^{Hj+N-1}\!\!
% \left(w_i\tilde{s}_i - w_i\tilde{y}_i\right)^{2}
% \right]
% \end{split}
% \label{eq:segmental_mse}
% \end{align}
\begin{align}
\begin{split}
\mathcal{E}_\text{DP}\distx{\tilde{\bm{y}}}{\tilde{\bm{s}}}
&= \distDP(\tilde{\bm{y}},\tilde{\bm{s}}; \bm{p}) \\
&= - \frac{1}{J} \sum_{j=0}^{J\!-\!1} p_{j}
\left[ 10\log_{10}
\frac{
    \sum_{i=Hj}^{Hj+N-1}
    \left(
      {w}_{i-Hj}\tilde{s}_{i}
    \right)^{2}
}{
    \sum_{i=Hj}^{Hj+N-1}
    \left(
      {w}_{i-Hj}\tilde{r}_{i}
    \right)^{2}
}
% \sum_{i=Hj}^{Hj+N-1}\!\!
% \left(w_i\tilde{s}_i - w_i\tilde{y}_i\right)^{2}
% \log_{10}\left[ \frac{\sum_t (w_i\tilde{s}_{it})^{2} }{\sum_t (w_i\tilde{s}_{i} - w_i\tilde{y}_{i} )^{2} } \right]
\right]
\end{split}
\label{eq:segmental_snr_loss}
\end{align}}%
% For brevity, we notate the data purification loss function, negative weighted segmental SNR, as $\distDP$.
Here, $J$ is the number of frames $\ceil{\frac{L}{H}}$.
Additionally, the residual vector is defined as $\tilde{\bm{r}} = \tilde{\bm{s}} - \tilde{\bm{y}}$.
This regressive model $h$ does not need to have pinpoint accuracy; as shown in Fig. \ref{fig:snr_prediction}, as long as $\hat{\bm{\alpha}}$ decently approximates $\bm{\alpha}$, the weights $\bm{p}$ will accurately reflect the position of speech-prominent frames in the data. If we substitute $\mathcal{E}_\text{DP}$ for $\mathcal{E}$ into the original PseudoSE loss function---Eq. \ref{eq:loss_pseudose}---we obtain a new data purified loss function:
\begin{equation}
    \mathcal{L}_\text{PseudoSE+DP} = \mathcal{E}_\text{DP}\distx{\tilde{\bm{y}}}{\tilde{\bm{s}}}.
\label{eq:loss_pseudose_dp}
\end{equation}
\change{Note that the slope of the logistic function could be further controlled by using an additional temperature weight applied to $\hat{\bm{\alpha}}$, which we opt not to investigate to focus more on the main contributions.}

Though substituting $\mathcal{E}_\text{DP}$ within the PseudoSE loss function is straightforward, it requires more nuance with the CM loss function. CM utilizes pairwise inputs, so therefore, we must compute pairwise weights as well.
\change{
\begin{align}
\bp^\oplus = \sigma(h(\tilde{\bs}^\oplus)), \hspace{1ex}
\bp^\ominus_1 = \sigma(h(\tilde{\bs}^\ominus_1)), \hspace{1ex}
\bp^\ominus_2 = \sigma(h(\tilde{\bs}^\ominus_2))
\end{align}
}%
Specifically in the case of positive pairs, the underlying pseudo source is the same, which is why there is only a single set of weights $\bp^\oplus$. Negative pairs are made up of two pseudo sources, so there are two sets of weights. For the negative pair estimate-to-estimate losses, we use the product of the two weight vectors, i.e. $\bp^\ominus = \bm{p}_1^\ominus \cdot \bm{p}_2^\ominus$. Using the appropriate weights for every term, we rewrite Eq. \eqref{eq:loss_positive} and Eq. \eqref{eq:loss_negative} as:
\change{
\begin{align}
\begin{split}
    \mathcal{L}_{p\text{+DP}} =
    & \distDP(\tilde{\bm{y}}_1^\oplus,\tilde{\bm{s}^\oplus}; \bm{p}^\oplus) \hspace{1mm}+ \\
    & \distDP(\tilde{\bm{y}}_2^\oplus,\tilde{\bm{s}^\oplus}; \bm{p}^\oplus) \hspace{1mm}+ \\
    & \lambda_p \big[ \distDP(\tilde{\bm{y}}_1^\oplus, \tilde{\bm{y}}_2^\oplus; \bm{p}^\oplus) \big]
\end{split}
\label{eq:loss_positive_dp} \\[2ex]
\begin{split}
    \mathcal{L}_{n\text{+DP}} =
    & \distDP(\tilde{\bm{y}}_1^\ominus,\tilde{\bm{s}_1^\ominus}; \bm{p}_1^\ominus) \hspace{1mm}+ \\
    & \distDP(\tilde{\bm{y}}_2^\ominus,\tilde{\bm{s}_2^\ominus}; \bm{p}_2^\ominus) \hspace{1mm}+ \\
    & \lambda_n \big[ \operatorname{max} \big( \distDP(\tilde{\bm{s}}_1^\ominus, \tilde{\bm{s}}_2^\ominus; \bp^\ominus), \\
    & \hspace{4.2em} \distDP(\tilde{\bm{y}}_1^\ominus, \tilde{\bm{y}}_2^\ominus; \bp^\ominus)
    \big) \big]
\end{split}
\label{eq:loss_negative_dp}
\end{align}
}%
The data-purified positive and negative loss functions may now be substituted in Eq. \eqref{eq:loss_contrastive} to obtain the overall CM+DP loss function:
\begin{equation}
\mathcal{L}_\text{CM+DP} = \sum_{t=1}^T\mathcal{L}_{p\text{+DP}}(t) + \sum_{t=1}^T\mathcal{L}_{n\text{+DP}}(t).
\label{eq:loss_contrastive_dp}
\end{equation}

\section{Experiment Setup}
\label{sec:experiment_setup}

In our experiments, we compare the baseline fully-supervised approach with the two proposed self-supervised approaches for training a personalized speech enhancement model. 
\change{Note that there are two rounds of model training (Fig. \ref{fig:overview}): one round that pretrains the model, and another ``finetuning'' round that only uses the available clean target speaker data (either \SI{5}{\sec} or \SI{30}{\sec}).}
We also assess the benefits of adding the data purification step to both self-supervised methods. We use the following shorthand notation to refer to each pretraining method:

\begin{itemize}
    \item \textbf{SE}: Models trained to minimize Eq. \eqref{eq:loss_baseline}. This is our generalist baseline, the speaker-agnostic speech enhancement system. It generalizes well only if its model capacity is large enough. % Otherwise, it could be still used as a pretraining method, followed by FSL using speaker-specific data if available.
    
    \item \textbf{PseudoSE}: Models trained to minimize Eq. \eqref{eq:loss_pseudose}. The proposed self-supervised method relies solely on noisy speaker-specific data $\SSnpretrain$. 
    
    \item \textbf{PseudoSE+DP}: Models trained to minimize Eq. \eqref{eq:loss_pseudose_dp}. This method refines the prior method through data purification. That is, the model uses a weighted segmental MSE as its discrepancy function in order to filter out noise-dominant frames within $\SSnpretrain$.
    
    \item \textbf{CM}: Models trained to minimize Eq. \eqref{eq:loss_contrastive}. This self-supervised method uses pairwise inputs that share either the same pseudo source or injection noise. CM provides additional regularization to PseudoSE through the contrastive loss terms.
    % Instead of random initialization, a model is first trained to minimize Eq. \eqref{eq:se_objective}, then finetuned to minimize Eq. \eqref{eq:pse_objective}.
    
    \item \textbf{CM+DP}: Models trained to minimize Eq. \eqref{eq:loss_contrastive_dp}. The pairwise weights inform the model of the mutual speech-dominant frames, thereby focusing the contrastive regularization specifically wherever the test-time speech is prominent.
\end{itemize}

\subsection{Datasets}

% Figure \ref{fig:methods} illustrates the differences between the baseline fully-supervised method, the two self-supervised methods, and the data purification process.
Table \ref{tab:corpora} provides a glossary of all the datasets and their notation used throughout this paper. Note that we subscript all datasets with either `tr', `vl', or `te' to indicate training, validation, or test partitions respectively.
% Table \ref{tab:equations} provides an overview of all the pretraining and finetuning procedures for personalized speech enhancement models.
For this paper, we limit the scope of personalization specifically regarding the test-time speaker and not the test-time environment. The extension of our methods towards environment adaptation is straightforward.

\begin{table*}[!htb]
\newcolumntype{Q}[1]{>{\raggedleft\let\newline\\\arraybackslash\hspace{0pt}}m{#1}}
    \centering
    \caption{Glossary of datasets paired with experiment-specific corpora.}
    \label{tab:corpora}
    \renewcommand{\arraystretch}{1}
    \begin{tabular}[t]{llrrm{3.3in}Q{1in}}\toprule
         \textbf{Set}
        & \textbf{Subset}
        & \textbf{Duration}
        & \textbf{Quantity}
        & \textbf{Description}
        & \textbf{Corpus}
        \\

        \midrule
        % \makecell[l]{\includegraphics[scale=0.5]{figures/icon_sa.pdf}}
         \multirow{2}{*}{$\SAfull$}
        & $\SAtrain$
        & \SI{443}{\hour}
        & {1,132} spkrs
        & \multirow{2}{3.3in}{Clean speech from many anonymous speakers}
        & \multirow{2}{*}{\makecell[r]{LibriSpeech \cite{PanayotovV2015Librispeech}}}
        \\
        & $\SAval$
        & \SI{8}{\hour}
        & \num{20} spkrs
        & 
        &
        \\
        \cmidrule{1-6}
        % \makecell[l]{\includegraphics[scale=0.5]{figures/icon_ss.pdf}}
         \multirow{7}{*}{$\SSfull$}
        & $\SSpretrain$
        & \SI{22.5}{\min\per\spkr}        
        & \multirow{7}{*}{\num{20} spkrs}
        & \multirow{2}{3.3in}{Target speaker's noisy speech (corrupted by $\Mfull$), referred to as the set of \textit{premixture} data}
        & \multirow{7}{*}{\makecell[r]{LibriSpeech \cite{PanayotovV2015Librispeech}}}
        \\
        & $\SSpreval$
        & \SI{30}{\sec\per\spkr}        
        &
        &
        &
        \\
        \cmidrule{2-3}\cmidrule{5-5}
        & $\SStrain$
        & \SI{60}{\sec\per\spkr}
        &
        & \multirow{2}{3.3in}{Target speaker's provided clean speech (only available in FSL context)}
        &
        \\
        & $\SSval$
        & \SI{30}{\sec\per\spkr}
        &
        &
        &
        \\
        \cmidrule{2-3}\cmidrule{5-5}
        & $\SStest$
        & \SI{30}{\sec\per\spkr}
        &
        & Set-aside clean speech from the target speaker used only for objective model evaluation
        &
        \\
        \cmidrule{1-6}
        % % \makecell[l]{\includegraphics[scale=0.5]{figures/icon_m.pdf}}
         \multirow{2}{*}{$\Mfull$}
        & $\Mtrain$
        & \SI{48}{\hour}
        & {13,339} noises
        & \multirow{2}{3.3in}{Premixture noises corrupting the majority of target speaker's utterances; on their own, inaccessible
        %to the service provider for
        during model training}
        & \multirow{2}{*}{\makecell[r]{FSD50K \cite{FonsecaE2020fsd50k}}}
        \\
        & $\Mval$
        & \SI{7}{\hour}
        & {1,929} noises
        &
        &
        \\
        \cmidrule{1-6}
        % % \makecell[l]{\includegraphics[scale=0.5]{figures/icon_n.pdf}}
         \multirow{4}{*}{$\Nfull$}
        & $\Ntrain$
        & \SI{5}{\hour}
        & \num{616} noises
        & \multirow{2}{3.3in}{Injection noises used
        %by service provider for 
        during model pretraining and fine-tuning}
        & \multirow{4}{*}{\makecell[r]{MUSAN \cite{SnyderD2015MUSAN}}}
        \\
        & $\Nval$
        & \SI{0.5}{\hour}
        & \num{60} noises
        &
        &
        \\
        \cmidrule{2-5}
        & $\Ntest$
        & \SI{0.5}{\hour}
        & \num{60} noises
        & Injection noises never seen during any model training, used to prepare target speaker-specific test sets
        &
        \\
        \bottomrule
    \end{tabular}
\end{table*}

In order to report objective signal improvement results, we designed experiments that simulate the personalization context. We therefore artificially mix signals from three publicly-available audio datasets: we use LibriSpeech \cite{PanayotovV2015Librispeech} for clean speech recordings, FSD50K \cite{FonsecaE2020fsd50k} for premixture noises, and MUSAN \cite{SnyderD2015MUSAN} for additionally injected noises. 

Out of the LibriSpeech \textit{train-clean-100} subset, we set aside \num{20} speakers to be the personalization targets; in other words, there are $K=20$ speaker-specific datasets $\mathbb{S}^{(k)}$ where $k\in\{1,\ldots, K\}$. We omit the speaker index $k$ going forward to simplify notation. The remaining speakers within Librispeech's \textit{train-clean-100} and \textit{train-clean-360} subsets are consolidated into the speaker-agnostic dataset $\mathbb{G}$. For all speech and noise corpora, we discard audio files shorter than \SI{4}{\sec} and resample everything to \SI{16}{\kilo\hertz}.

We partition each speaker-specific dataset $\mathbb{S}$ into five sets as shown in Table \ref{tab:corpora}. The utterances are sorted by duration and grouped such that approximately \SI{30}{\sec} are available for testing the model ($\SStest$), \SI{30}{\sec} for validating finetuned models ($\SSval$), \SI{60}{\sec} for FSL-based finetuning ($\SStrain$), and \SI{30}{\sec} to validate the self-supervised pretraining methods ($\SSpreval$). The remaining \SI{22.5}{\min} are used for pretraining ($\SSpretrain$). Subsequently, for each of the \num{20} personalization targets, a test set is constructed using \num{100} mixtures that combine $\SStest$ with $\Ntest$.

$\Mtrain$ and $\Mval$ follow the train and val splits provided in FSD50K's \textit{dev} folder. Using the FSD50K provided tags, we omit files tagged as either ``speech'' or ``music''. 

The unseen test-time noises, $\Ntest$, are derived from MUSAN's \textit{sound-bible} folder. Using MUSAN's \textit{free-sound} folder, sixty random noises are set aside for $\Nval$ and the remaining noises make up $\Ntrain$.

These datasets are carefully chosen and arranged to represent our use-case scenarios. First, we need a large dataset  $\SAfull$ to encompass diverse speaker characteristics. Second, we ensure that the \num{20} personalization target speakers have enough clean speech signals $\SSpretrain$ in order to simulate the abundant premixture signals $\SSnpretrain$. The premixture noise sources $\Mtrain$ are also very diverse so as to simulate various acoustic environment the user can be situated in. Tallying the unique FSD50K audio tags, our experiment simulates each of the \num{20} target speakers being degraded by approximately \num{160} noise types. Through the premixture process, we combine $\bm{s}$ and $\bm{m}$ such that the SNR is uniformly random between \SIrange{0}{15}{\decibel}. Psychoacoustic research has shown that this SNR range describes many real-world sound environments \cite{OlsenW1998realsnrs, SmedsK2015realsnrs}. Lastly, mixtures, which are made using the injection noise set $\Nfull$, have SNRs chosen uniformly at random between \SIrange{-5}{5}{\decibel}.

There are other choices of speech datasets, besides Librispeech, which contain real-world recordings of in-the-wild noisy speech, e.g., AudioSet \cite{GemmekeJ2017audioset}. Although our proposed self-supervised training methods are intended for in-the-wild data, it is often the case that such datasets do not possess enough noisy recordings from a single consistent speaker. More importantly, in order for us to report objective signal improvement, we require ground-truth clean speech recordings from the test-time speaker. Therefore, our experiments simulate the personalization problem through the three separate corpora, constructing numerous artificial mixtures and premixtures.

\subsection{Metrics}
\label{sec:metrics}

% There have been many different discrepancy metrics studied as potential loss functions for monaural speech enhancement algorithms \cite{KolbaekM2020ieeeacmaslp}. Minimizing time-domain mean-square error (MSE) does correlate well with improving the perceptual signal quality. 

% There are many potential choices for the loss function $\mathcal{E}$ . For this paper, we set $\mathcal{E}$ to the time-domain mean square error ($\mathcal{L}_{\text{MSE}}$), which is the per-sample squared distance. For estimate signals $\hat{\bm{v}}$ and their respective target signals $\bm{v}$ with length $L$, MSE is defined as:

% \begin{equation}
% \mathcal{L}_{\text{MSE}}(\hat{\bm{v}}, \bm{v}) = \frac{1}{L} \sum_{i=0}^L \left(\bm{v}_i - \hat{\bm{v}}_i \right)^{2}
% \label{eq:mse}
% \end{equation}

% Central to all of the loss functions for the proposed pretraining methods ($\mathcal{L}_\text{SE}$, $\mathcal{L}_\text{PseudoSE}$, and $\mathcal{L}_\text{CM}$) is the discrepancy function: $\mathcal{E}$ or $\mathcal{E}_\text{DP}$. 
% \change{As mentioned in \ref{sec:generalist}, we opt to use SI-SDR as the discrepancy function for all our experiments.}
\change{
With our experiments, we report three metrics frequently used in speech enhancement research: SDR \cite{VincentE2006ieeeaslp}, PESQ \cite{RixA2001pesq}, and extended STOI \cite{TaalC2010icassp}.
Unlike the objective measurement SDR, the latter two are perceptual metrics that highly correlate to speech intelligibility.
As all of our loss functions are SDR-based, our models in this experiment do not explicitly optimize for intelligibility.
Each one of the \num{20} target speakers has their own test set, made up of \num{100} mixtures with input SNR between \SIrange{-5}{5}{\decibel}. All three metrics are computed between the estimate signals and their corresponding target signals.
}

\begin{table}[t!]
\centering

\caption{List of model architectures, configurations, and sizes.}
\renewcommand{\arraystretch}{1.2}
\begin{tabularx}{\columnwidth}{
    lll
    S[table-number-alignment=right, table-format=3.1, round-mode=places, round-precision=1]
    S[table-number-alignment=right, table-format=3.1, round-mode=places, round-precision=1]}
\toprule
\textbf{Architecture}
& \textbf{Size}
& \textbf{Configuration}
& \textbf{\makecell[r]{Params}}
& \textbf{\makecell[r]{MACs}} \\
% & \textbf{\makecell[r]{Disk Space\\(\mebi\byte)}} \\
\midrule
% \multirow{4}{*}{\makecell[c]{GRUNet}}
% & Large & $H_s=256$
% & \prefix{1118721} & \prefix{281569792} \\
% & Medium & $H_s=128$
% & \prefix{412161} & \prefix{103773440} \\
% & Small & $H_s=64$
% & \prefix{169473} & \prefix{42633856} \\
% & Tiny & $H_s=32$
% & \prefix{75777} & \prefix{19003712} \\
% \midrule
\multirow{4}{*}{\makecell[c]{Conv-TasNet}}
& Large & \makecell[l]{$B_c=64$, $H_c=256$}
& \prefix{1037021} & \prefix{8353259712} \\
& Medium & \makecell[l]{$B_c=32$, $H_c=128$}
& \prefix{437757} & \prefix{3531078560} \\
& Small & \makecell[l]{$B_c=16$, $H_c=64$}
& \prefix{224141} & \prefix{1808029968} \\
& Tiny & \makecell[l]{$B_c=8$, $H_c=32$}
& \prefix{138837} & \prefix{1118516168} \\
% \midrule
% \multirow{4}{*}{\makecell[c]{DPTNet}}
% & Large & \makecell[l]{$H_s=144$}
% & \prefix{1054145} & \prefix{136080439680} \\
% & Medium & \makecell[l]{$H_s=72$}
% & \prefix{367553} & \prefix{47261239680} \\
% & Small & \makecell[l]{$H_s=36$}
% & \prefix{148673} & \prefix{18839095680} \\
% & Tiny & \makecell[l]{$H_s=18$}
% & \prefix{70337} & \prefix{8624887680} \\
\bottomrule
\end{tabularx}
\label{tab:experiment_configurations}
\end{table}

\subsection{Neural Network Architectures}

% We assess the performance of generalist and specialist speech enhancement models across two neural network architectures, each of which has a small, medium, and large-sized variant. As shown in Table \ref{tab:experiment_configurations}, we designed these model configurations such that the total number of trainable parameters is near-equal with each model size.

% The first architecture we experiment with is a two-layer gated recurrent unit network (GRU) \cite{ChoK2014emnlp}. It performs monaural speech enhancement via time-frequency masking \cite{NarayananA2013icassp}: first, input waveforms are converted to the time-frequency domain using the short-time Fourier Transform (STFT). For the STFT, we use a frame size $N=1024$ and a hop size $H=256$. Because our model inputs are \SI{4}{\sec} clips sampled at \SI{16}{\kilo\hertz}, the resulting spectrograms have \num{513} bins and $J=250$ frames. The GRU network then estimates a binary weighting matrix (mask) out of the noisy magnitude spectrogram. Their element-wise product produces a denoised spectrogram, which we convert back to the time-domain using the inverse STFT. The three size variants modify the hidden size of both GRU layers ($H_s$).

Well-established neural network approaches for speech enhancement utilize time-frequency masking.
In order to overcome latency and phase reconstruction limitations, more recent neural network algorithms operate in an end-to-end manner, i.e., by learning a mapping directly between the time-domain input and output signals \cite{YoshiiK2013ismir, VenkataramaniS2018asilomar, StollerD2018waveunet}.
To that end, we assess the performance of generalist and specialist speech enhancement models using ConvTasNet (CTN), which is a popular fully-convolutional time-domain model for audio separation \cite{LuoY2019conv-tasnet}. It operates as follows: first, the encoder module maps input waveforms into latent representations. Then, the separation module calculates a multiplicative mask that separates the target source. Lastly, the decoder module maps the masked latent features back to the time-domain, yielding estimate waveforms. The CTN architecture may be generalized to separate multiple audio sources; however, our separation module estimates only one mask to specifically separate speech from noise. With each size variant, we adjust the number of channels in the separation module's bottleneck ($B_c$) as well as the number of channels in convolutional blocks ($H_c$) such that the expansion ratio $H_c/B_c \approx 4$ \cite{SandlerM2018mobilenetsv2}.

As shown in Table \ref{tab:experiment_configurations}, we designed a tiny, small, medium, and large-sized variant of CTN such that the total number of trainable parameters is less than or equal to one million. \change{MACs indicate the number of multiply-accumulate operations, correlating to computational complexity}. As shown in prior work, personalized speech enhancement is a subset of the broader universal speech enhancement problem, therefore specialist models can achieve comparable performance to generalist models using fewer parameters \cite{SivaramanA2020interspeech, SivaramanA2021interspeech}. Through our experiments, we report the performance of the different sized variants to observe whether this model compression trend applies to the modern fully-convolutional models. 

% Our second architecture for this experiment is Conv-TasNet (CTN), a fully-convolutional time-domain audio separation network that performs end-to-end source separation \cite{LuoY2019conv-tasnet}. It operates as follows: first, the encoder module maps input waveforms into high-dimensional latent representations. Then, the separation module calculates a multiplicative mask that separates the target source. Lastly, the decoder module maps the masked latent features back to the time-domain, yielding estimate waveforms. The CTN architecture may be generalized to separate multiple audio sources; however, our separation module estimates only one mask to separate speech from noise specifically. With each size variant, we adjust the number of channels in the separation module's bottleneck ($B_c$) as well as the number of channels in convolutional blocks ($H_c$) such that the expansion ratio $H_c/B_c \approx 4$ \cite{SandlerM2018mobilenetsv2}.

\subsection{Implementation Details}

\change{All models were implemented using PyTorch \cite{PaszkeA2019pytorch} and trained on NVIDIA Tesla V100 graphics cards.}
We used the ConvTasNet implementation found in the Asteroid package \cite{ParienteM2020asteroid}.
All experiments have a fixed batch size of \num{64}.
We utilize the Adam optimizer \cite{KingmaD2015adam} with an initial learning rate of \num{1e-3}.
\change{When finetuning over clean speech data ($\SStrain$), the learning rate is instead \num{1e-4}.}
For every \num{1000} mixtures processed, we compute SDR improvement averaged over a fixed set of \num{100} validation mixtures; the trial is terminated if the mean validation SDR does not improve after \num{100000} further mixtures.

\change{Using the described early stopping scheme, we observed various trends with regards to the training time.
On average, generalist models trained over \SI{1.4}{\mega\nothing} mixtures for all four sizes, whereas specialist models trained over \SI{851}{\kilo\nothing}, \SI{803}{\kilo\nothing}, \SI{637}{\kilo\nothing}, and \SI{593}{\kilo\nothing} mixtures for the Tiny, Small, Medium, and Large model sizes respectively.
When these models undergo finetuning using \SI{5}{\sec} of clean speech, the specialists converge after seeing \SI{6.4}{\kilo\nothing}, \SI{6.0}{\kilo\nothing}, \SI{5.7}{\kilo\nothing}, and \SI{5.2}{\kilo\nothing} mixtures for the Tiny, Small, Medium, and Large model sizes respectively.}

Source code for this experiment may be found at \url{https://saige.sice.indiana.edu/research-projects/pse-ssl-dp}.

% \section{Experimental Results and Discussions}
\section{Experiment Results}

\subsection{Contrastive Mixtures Ablation Study}
\label{sec:ablation}

Prior to starting the full personalization experiment, we first determine optimal values for $\lambda_p$ and $\lambda_n$ which modulate the contrastive mixtures positive and negative loss terms---Eq. \eqref{eq:loss_positive} and \eqref{eq:loss_negative} with DP variants \eqref{eq:loss_positive_dp} and \eqref{eq:loss_negative_dp}.
Therefore, we run an ablation study of contrastive mixtures by performing a grid search over potential choices: \num{1}, \num{1e-1}, \num{1e-2}, \num{1e-3}, \num{1e-4}, and \num{0}.
We can assess the effectiveness of the positive and negative pairs by setting either one of $\lambda_n$ to $\lambda_p$ to \num{0}, respectively.
For the purposes of the ablation study, we run experiments in which the personalized speech enhancement system is fixed as a small ConvTasNet as specified in Table \ref{tab:experiment_configurations}.
This is done for three out of the twenty personalization target speakers from LibriSpeech.
This results in \num{216} total trials, given that there are \num{36} $\lambda$ combinations and \num{3} target speakers, plus the option for data purification to be enabled or disabled.
We report the validation set signals' SDRs after pseudo-enhancement, averaged across the three speakers and across \num{100} validation premixtures utterances.
In summary, a small ConvTasNet is trained over speaker-specific premixtures using a batch size of \num{64}, a learning rate of \num{1e-3}, and the \textbf{CM} loss function: either Eq. \eqref{eq:loss_contrastive} or \eqref{eq:loss_contrastive_dp}.

\begin{figure}
    \centering
    \includegraphics[width=\columnwidth]{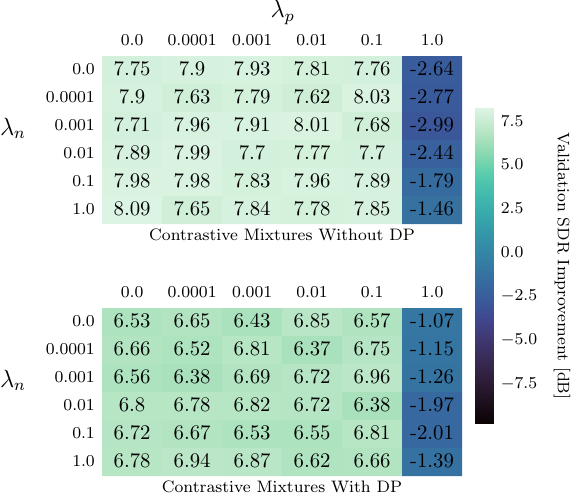}
    \caption{Ablation study of the contrastive mixtures (CM) loss function, where we vary $\lambda_p$ and $\lambda_n$ to adjust the contribution of the positive and negative pair loss terms. Pseudo-enhancement is performed using the small ConvTasNet architecture, and results are averaged across three test-time speakers.}
    \label{fig:ablation_study}
\end{figure}

From Fig. \ref{fig:ablation_study}, we observe that there are many working combinations of $\lambda_p$ and $\lambda_n$, so long as $\lambda_p < 1$.
This suggests that CM is robust to the hyperparameter selection.
The top-left corner of both subplots represents models trained with the contrastive loss terms disabled---effectively, trained through PseudoSE.
By scanning the left-most column and top-most row, we can see that the negative pair loss terms improve the model more significantly than the positive pair loss terms.

When pretraining without data purification, the most-optimal configuration happens to be with $\lambda_n = 1$ and $\lambda_p = 0$, yielding a \SI{0.34}{\decibel} (or \SI{4.4}{\percent}) improvement over PseudoSE.
If both $\lambda$s are non-zero, we see slight variations in the validation performance.
When the noisy training data is non-purified, it is possible that the positive pair contrastive loss compels the model to enforce similarity on highly degraded pseudo-sources.
These cases emphasizing premixture noise reconstruction similarity could cause the learned parameters to drift slightly away from speech-focused personalization.

The bottom subplot of Fig. \ref{fig:ablation_study} shows models pretraining through CM with data purification.
Here, the most-optimal configuration is $\lambda_n = 0.001$ and $\lambda_p = 0.1$; the self-supervised model sees a \SI{0.43}{\decibel} (or \SI{6.6}{\percent}) improvement over PseudoSE.
Notably, the positive pair-only models are able to obtain a \SI{0.32}{\decibel} (or \SI{4.9}{\percent}) improvement.
With the CM loss functions weighted towards speech-dominant frames, we see that the positive and negative loss terms synergies more effectively.

One last observation is that the validation SDR of models using DP is overall lesser than that of models not using DP.
This follows our hypothesis that the DP-based loss functions are more similar to the true fully-supervised speech enhancement loss.
Note that all the self-supervised models are assessed on pseudo enhancement during validation.
Therefore, it is understandable that the DP-based models have a lesser validation SDR improvement.
The metrics computed at test-time assess true speech enhancement performance; therefore, observing this trend during validation alludes to greater enhancement.

Given our observation that CM works for many configurations, as a convenience for all other experiments, we set $\lambda_n = 0.1$ and $\lambda_p = 0.1$ with both non-purified and purified models. 

\subsection{Efficient Personalized Speech Enhancement Study}

\begin{figure*}[t]
    \centering
    \includegraphics[width=\textwidth]{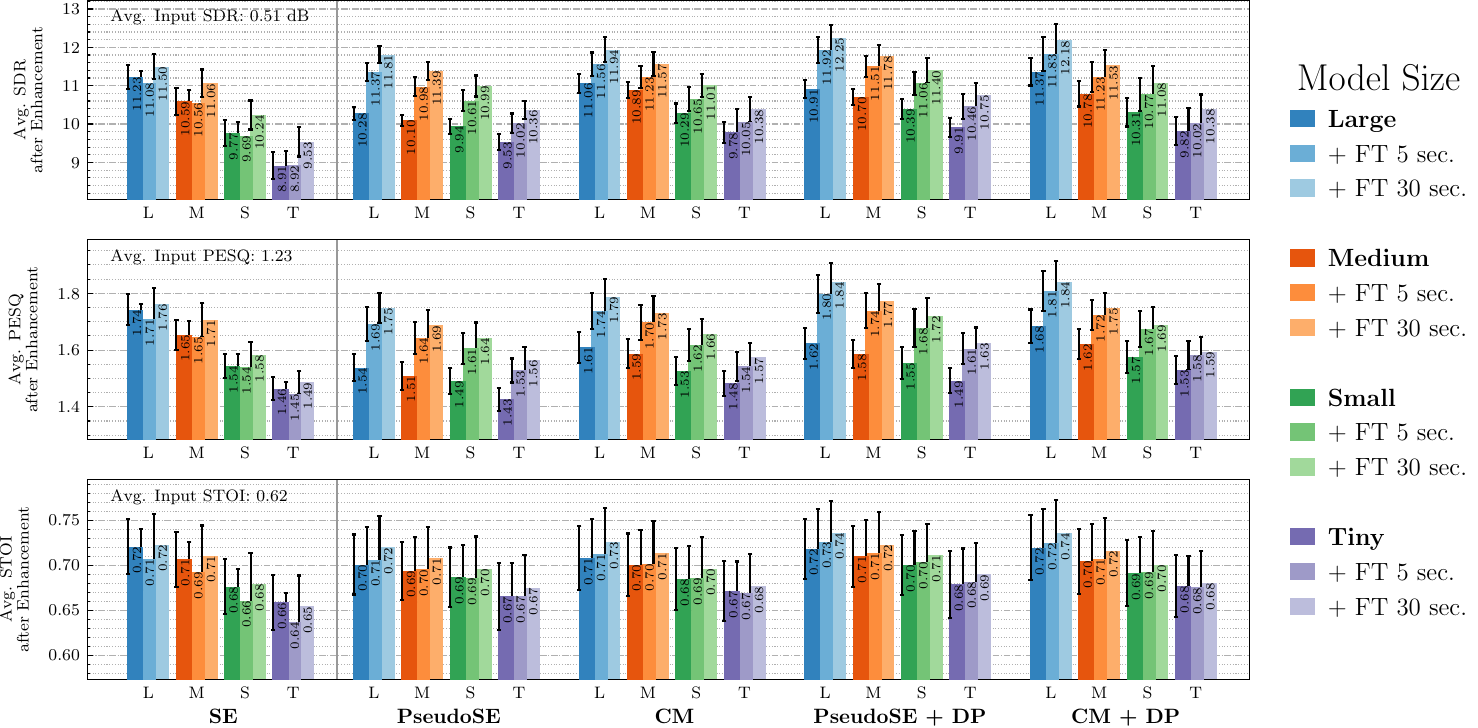}
    \caption{
    Experiment results. The improvement for each metric (SDR, PESQ, and STOI) may be calculated by subtracting the average input value from the average value after enhancement.
    The plotted metrics are averaged \change{over the 100 test set utterances} for each of the \num{20} target speakers.
    \change{The shading of each bar corresponds to the amount of clean speech data from the target speaker used for finetuning}: \SI{0}{\sec} (i.e., no finetuning), \SI{5}{\sec}, and \SI{30}{\sec}. 
    % Performances reported using \SI{0}{\sec} represent the ZSL capabilities of the pretraining method. We explore FSL contexts of \SI{5}{\sec} and \SI{30}{\sec} to investigate high and low amounts of data efficiency. The left-most boxplots within the SE column represent unpersonalized / generalist performance.
    \change{Error bars show the specific 95\%-confidence interval per model and training configuration, averaged over all target speakers.}
    }
    \label{fig:results}
\end{figure*}

Next, we discuss the results from the main experiment. As described in in Sec. \ref{sec:experiment_setup}, we consider \num{20} target speakers, \num{4} model sizes, \num{4} self-supervised pretraining methods, and \num{2} possible amounts of clean speech data. In terms of model checkpoints, there are \num{4} unadapted SE models, \num{160} fine-tuned SE models, \num{320} self-supervised PSE models, and \num{640} fine-tuned PSE models, resulting in a total of \num{1124} trials.

Figure \ref{fig:results} shows test set results in terms of the three signal quality metrics listed in Sec. \ref{sec:metrics}. Results are averaged across \num{20} test-time speakers, with different bars representing different model sizes and training configurations. 

% Figure \ref{fig:finetuning} shows the impact of finetuning on clean speech with regards to the different initialization schemes, with subplot columns for individual test-time speaker and rows for each model architecture.

\subsection{ZSL Personalization Performance}

Bars with the darkest shading represent the performance of models in the ZSL personalization context, in which the models lack access to clean speech from the target speaker.
% The left-most boxes (in green) in each cell of Figure \ref{fig:results} indicate the performance of the models before applying any finetuning using the ground-truth speech signals $\SStrain$. We labeled models trained for the ZSL context on the x-axis as \SI{0}{\sec} as they do not utilize any labelled data. 

\subsubsection{Generalist Models' Performance}
The \textbf{SE} column's left-most bars show the performance of the bare generalist models' performance.
The generalists are able to enhance the noisy test-time speakers in all cases, but it is clear that the larger models (bars labeled L or M) show much better generalization performance (up to \num{11.23}{\decibel} SDR after enhancement) than the smaller ones (lower rows). 
For the tiny generalist models, the average SDR after enhancement is \num{8.92}{\decibel}.
This \SI{2.31}{\decibel} range reinforces our argument that the smaller generalists tend to be poorer in generalization.
Note that these baseline SE models are non-personalized. 
As they are without any adaptation, we can observe that the generalists' performance correlates with the architectural complexity because they are all trained using a large dataset. 

\subsubsection{Personalization using PseudoSE}

The \textbf{PseudoSE} column shows the performance of the self-supervised models trained through pseudo enhancement of noisy speech targets.
The model inputs are doubly-degraded observations of the test-time speaker ($\SSnpretrain$ is mixed with additional noise sources $\Ntrain$), and the model na\"ively recovers the pseudo-source.
There is a chance that the pseudo targets are too far from clean speech, deviating the learned parametric function from the ideal personalized SE model.
However, it is also possible that some parts of these pseudo speech sources are somewhat clean enough in order for the model to learn the target speaker's speech traits.
The left-most bars (darkest shade) of the \textbf{PseudoSE} column do reveal success in personalization---note that the confidence interval of SDR enhancement narrows by using PseudoSE pretraining compared to SE pretraining.
This trend is less obvious with perceptual metrics PESQ and STOI, but it is to be expected as the models' loss functions are SDR-based.
PseudoSE does produce improvements over the \textbf{SE} pretraining when the models are tiny (\num{9.53} vs. \num{8.91}) or small (\num{9.94} vs. \num{9.77}).
However, when the model complexity is large enough, we see that PseudoSE is unable to compete with the generalist model.
Compare the largest model trained using PseudoSE against the largest speaker-agnostic SE model (\num{10.28} vs. \num{11.23}).
Therefore, we conclude that PseudoSE's personalization performance is significant only when the model is incapable of learning from the large generic dataset.

\subsubsection{Impact of DP with PseudoSE}

As shown in our prior work \cite{SivaramanA2021interspeech}, DP can identify cleaner frames from premixture signals $\SSnpretrain$ and improve the usability of the target speaker's noisy speech signals.
We observe a similar trend with our ConvTasNet-based experiments.
In particular, our results show that the \textbf{PseudoSE+DP} pretraining scheme in the ZSL context yields greater improvements over the plain \textbf{PseudoSE} in the large model than in the smaller ones.
For example, introducing DP lifts the average performance of PseudoSE by \SI{0.63}{\decibel} (\num{10.91} vs. \num{10.28}) in the large models, while the tiny models only see an average boost of about \SI{0.38}{\decibel} (\num{9.91} vs. \num{9.53}).
Because PseudoSE's efficacy is limited in the large models, the gains from introducing data purification are more prominent.
However, it is still the case that the tiny model gains the most from the consolidated personalization process, e.g., a \num{1.0}{\decibel} improvement from the baseline SE model (\num{9.91} vs. \num{8.91}). 

\subsubsection{Personalization using CM}

The ZSL results of the \textbf{CM} column are noteworthy because they compete with the \textbf{PseudoSE+DP} results despite using non-purified data.
For example, \textbf{CM} results in better performance than \textbf{PseudoSE+DP} in large models (\num{11.06} vs. \num{10.91}) and works on par with \textbf{PseudoSE+DP} in small or tiny models.
This shows that the proposed CM loss functions help the model learn robust features for personalized SE even though the signals used are noisy observations (or unlabeled in the sense of classification).
These results validate the powerful feature learning capabilities of contrastive learning.
Although the contrastive self-supervised learning paradigm has been explored in other research areas (e.g., SimCLR for computer vision), we note that the proposed CM pretraining method is specifically designed for source separation problems. 

\subsubsection{Impact of DP on CM}

We find that \textbf{CM+DP} does not introduce significant improvements except with the largest model.
This is likely due to the robust feature learning ability of CM, which is already competitive with the DP process. 

\subsubsection{Model Compression}

Among the tiny-sized models, the best-performing ZSL method for personalization is \textbf{PseudoSE+DP} which produced an average SDR improvement of \num{9.91}{\decibel}.
We see that the personalized tiny model outperforms the generalist small model (\num{9.77}{\decibel}), although it uses \num{62}{\percent} fewer model parameters and multiply-accumulate operations (MACs) according to Table \ref{tab:experiment_configurations}.
Likewise, the personalized small model comes within striking distance the medium-sized generalist ($10.39$ vs. $10.59$) using less than \num{52}{\percent} of the spatial and computational complexity.
Finally, the best medium model after the \textbf{CM} personalization (\num{10.89}\decibel) has its confidence interval overlapped with that of the largest SE baseline (\num{11.23}\decibel), although its model complexity is less than $44\%$.
From this we can conclude that, for lower-complexity models, the proposed self-supervised ZSL personalization may be viewed as a lossless model compression paradigm. 

% \subsubsection{\todo{Personalization can reduce the variance of the models (height of the boxes), meaning there's no big failure cases. Flesh out from this claim.}}
\subsubsection{Success of Personalization}

The height of the error bars indicate the 95\%-confidence interval of each model and training configuration seen across the \num{20} target speakers.
Using \textbf{SE} generalist pretraining, we observe that this variance can be as much as \SI{0.9}{\decibel} for the tiny-sized models or \SI{0.7}{\decibel} with the large-sized models.
Through the proposed PseudoSE and CM methods, we see that the variance universally decreases in the ZSL context.
Therefore, our self-supervised pretraining methods successfully adapt to the nuances of each test-time speaker despite being trained using only noisy data.
Our results do show that introducing DP increases the variance in performance once again.
This is to be expected as the availability of near-clean frames can differ greatly between speakers.
Similarly, DP's reliance on the external SNR predictor model is also a contributing factor.

\subsection{FSL Personalization Performance}
% In the FSL context, the test-time speaker provides their clean speech ($\SStrain$) for finetuning.
Bars with lighter shading represent the FSL context, wherein models have \SI{5}{\sec} or \SI{30}{\sec} of clean speaker-specific data to finetune over.

\subsubsection{Generalist Models' Performance}

We observe that all four sizes of the baseline models pretrained as generalists (\textbf{SE}) are incapable of adapting over a small $\SStrain$ that has only \SI{5}{\sec} of data.
Using \SI{30}{\sec} of clean speech data does eventually produce gains for all model sizes.
The tiny-sized generalist sees the most significant gains (\SI{0.62}{\decibel}) whereas the large-sized generalist sees marginal benefit (\SI{0.27}{\decibel}). 
% \todo{Argue that FSL on \textbf{SE (Baseline)} isn't very effective when there's only 5 sec available. With 30 sec it's a bit better.}
This trend implies that the pretrained generalists are defined by model parameters that are too far from the ideally personalized counterpart, requiring much effort during the transfer learning process.
In other words, too few clean utterances do not suffice in achieving the domain adaptation.

\subsubsection{FSL after PseudoSE Initialization}

We reiterate that our self-supervised methods train using noisy speaker-specific data with premixture SNRs in the \SIrange{0}{15}{\decibel} range.
Hence, \textbf{PseudoSE} pretraining over this noisy data proves to be useful only for the tiny- and small-sized models (\num{9.53} vs. \num{8.91} and \num{9.94} vs. \num{9.77}), while the larger models do not benefit from the simple SSL setup.
However, with all model sizes, finetuning using only \SI{5}{\sec} of clean data results in a significant performance boost (%
\num{10.02} vs. \num{8.92}, %
\num{10.61} vs. \num{9.69}, %
\num{10.98} vs. \num{10.59}, and %
\num{11.37} vs. \num{11.08}).

Similar boosts also appear when using \textbf{PseudoSE+DP}, where all the performance scores are lifted by up to \SI{0.84}{\decibel} (\num{11.92} vs. \num{11.08} in the largest models).
Our results suggest that finetuning is much more effective due to the speaker-specific self-supervised pretraining.
By comparing the middle shaded bars in the \textbf{PseudoSE+DP} column with lightest shaded bars in the \textbf{SE} column, we can also see the data efficiency benefits of our self-supervised methods.
In particular, after the \textbf{PseudoSE+DP} pretraining, only \SI{5}{\sec} of clean speech for finetuning achieve a greater mean SDR improvement compared to generalists models finetuned using \SI{30}{\sec} of clean speech.
\textbf{PseudoSE+DP} achieves data efficiency with all model sizes (%
\num{10.46} vs. \num{9.53}, %
\num{11.06} vs. \num{10.24}, %
\num{11.51} vs. \num{11.06}, and %
\num{11.92} vs. \num{11.50}).
Our results show that through self-supervised pretraining, we are able to reduce reliance on the target speaker's private data by a factor of \num{6}.
% \todo{Argue that \textbf{PseudoSE} and \textbf{PseudoSE+DP} improves the FSL performance better; talk about the gap before and after applying 5sec FSL and then argue that they  are much better than the corresponding baseline performances. Emphasizing the performance of 5sec boxes is very important here, as it is related to the data efficiency argument.}

\subsubsection{FSL after CM Initialization}

In the ZSL context, \textbf{CM} pretraining produced notable improvements over \textbf{PseudoSE} likely due to the contrastive loss terms that introduce powerful regularization.
But we found that the performance gap between CM and PseudoSE is nearly negligible in the FSL context.
When it comes to data purification, we found that \textbf{CM+DP} was less effective in the FSL contexts than \textbf{PseudoSE+DP}.
This is perhaps due to the data purification learning objective being too different from the contrastive learning objective, leading to a slightly sub-optimal joint learning objective.
Nonetheless, for the ZSL scenario, CM pretraining without data purification has merit over PseudoSE, because it can alleviate the need for training a robust SNR predictor.
% \todo{I don't know what happened here but CM or CM+DP doesn't work better than PseudoSE or PseudoSE+DP as an initialization method---the improvement from FSL compared to the 0sec cases is not as good. You should still argue that CM or CM+DP is useful for the ZSL scenario, when you don't want to rely on the performance of an SNR predictor.}

\subsubsection{Model Compression}

Finetuning also augments the model compression benefits of personalization.
For example, we can use a small-sized \textbf{PseudoSE+DP} model finetuned with only \SI{5}{\sec} of clean speech to get \SI{11.06}{\decibel} SDR after enhancement on average.
This is on par with the largest \textbf{SE} model finetuned over the same amount of clean speech data (\SI{11.08}{\decibel}).
This example shows a lossless $78\%$ reduction in model parameters and MACs.

\section{Conclusion}
We put forward self-supervised learning approaches towards personalized speech enhancement, highlighting their ability to learn robust features from the target speaker's noisy observations. Our main ideas are based on the assumption that noisy utterances of the target speaker might be more available than clean speech. However, due to the noisy nature of those unlabeled data, we propose more sophisticated SSL treatments to learn useful features from them. PseudoSE sets up a pretext SE problem where the enhancement target is still a noisy utterance. In addition, data purification improves the usability of the unlabeled (thus noisy) speech signals by identifying cleaner frames and focus more on them. With the purification step, PseudoSE becomes more realistic. Contrastive mixtures add an additional regularization benefit to the loss function, so that the pretext task is more relevant to the original source separation problem. 

We observe that all these methods can act as a zero-shot personalization system which adapts to the target speaker's specificity with no additional clean speech used. In the few-shot learning context, we emphasize that the proposed SSL methods also serve as a better initialization scheme than a na\"ive generalist as the SSL methods learn from the target speaker's speech, even though it is contaminated. We found that the proposed systems quickly adapt using only a few seconds of test-user clean speech data, which is a too small amount for the baseline generalists to effectively perform transfer learning. Our results suggests that speaker-discriminative features can be found even in noisy recordings. The benefit of personalization is that it can reduce model complexity with no loss of SE performance, e.g., small personalized models perform as good as twice-larger general-purpose SE models. In addition, the proposed SSL methods make the few-shot learning-based personalization more data-efficient. Given that the transfer learning-based personalization requires clean speech data from the test-time users, reducing the required amount can improve the user experience. 

% Future Work? Should we mention that we tried DPRNNTasNet but could not find a working configuration below 1 million parameters?

\section*{Acknowledgments}
This material is based upon work supported by the National Science Foundation under Grant No. 2046963.

\bibliographystyle{IEEEtran}
\bibliography{mjkim.bib}

\section{Biography Section}
\vspace{-33pt}
\begin{IEEEbiography}
[{\includegraphics[width=1in,height=1.25in,clip,keepaspectratio]
{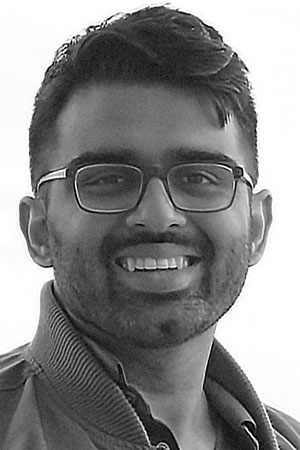}}]{Aswin Sivaraman}
is currently a Ph.D candidate at Indiana University in the Department of Intelligent Systems Engineering. He received his BS degree in electrical engineering from the University of Illinois at Urbana-Champaign in 2015. Aswin was previously with Qualcomm as a software engineer, and has been an intern for Google, X, Spotify, and Amazon. He is a recipient of the 2017 Graduate Fellowship at Indiana University and the Interspeech 2020 Student Travel Grant. His research focuses on deep learning methods for speech and music processing.
\end{IEEEbiography}

\vspace{-11pt}
\begin{IEEEbiography}
[{\includegraphics[width=1in,height=1.25in,clip,keepaspectratio]
{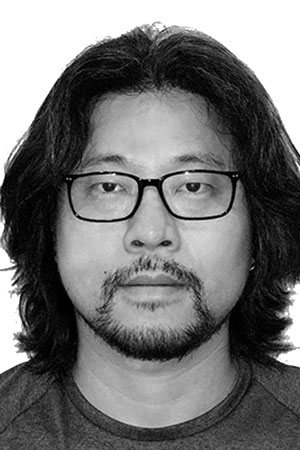}}]{Minje Kim}
(M’12–SM’19) is Assistant Professor of Intelligent Systems Engineering at Indiana University, where he is also affiliated with Data Science, Cognitive Science, and Statistics. He is also an Amazon Visiting Academic. He received the Ph.D. degree in Computer Science from the University of Illinois at Urbana-Champaign in 2016. Before that, he worked as a researcher at ETRI, Daejeon, Korea from 2006 to 2011. He is an IEEE Senior Member and a member of the IEEE AASP TC. He is a recipient of the NSF CAREER Award (2021), IEEE SPS Best Paper Award (2020), Google and Starkey’s grants for outstanding student papers at ICASSP 2013 and 2014, respectively. His research spans machine learning and audio signal processing.
\end{IEEEbiography}

\vfill
\end{document}